\documentclass[10pt, aps, prd, floats, floatfix, superscriptaddress, onecolumn,nofootinbib,amsmath,amssymb,longbibliography]{revtex4-2}

\usepackage{soul}
 \usepackage{enumerate}
 \usepackage{graphicx}
 \usepackage{multirow}
 \usepackage{xcolor}
 \usepackage{tablefootnote}
\usepackage{threeparttable}
 \usepackage{bm}
\usepackage{babel}
 \usepackage[normalem]{ulem}
 \usepackage{comment}
\usepackage[normalem]{ulem}
\usepackage{url}

\usepackage[acronym]{glossaries}
 \usepackage{physics}
\usepackage{comment}

\begin{document}

\title{A First Post-Friedmann Extension of the Schr\"odinger Approach to Cosmic Structure Formation}

\author{Yiwen Deng-Huang}
\affiliation{Dipartimento di Fisica e Astronomia Galileo Galilei, \\Universit\`a degli Studi di Padova, via Marzolo 8, I-35131, Padova, Italy}
\affiliation{INFN, Sezione di Padova, via Marzolo 8, I-35131, Padova, Italy}

\author{Daniele Bertacca}
\affiliation{Dipartimento di Fisica e Astronomia Galileo Galilei, \\Universit\`a degli Studi di Padova, via Marzolo 8, I-35131, Padova, Italy}
\affiliation{INFN, Sezione di Padova, via Marzolo 8, I-35131, Padova, Italy}
\affiliation{INAF- Osservatorio Astronomico di Padova, \\ Vicolo dell Osservatorio 5, I-35122 Padova, Italy}

\author{Sabino Matarrese}
\affiliation{Dipartimento di Fisica e Astronomia Galileo Galilei, \\Universit\`a degli Studi di Padova, via Marzolo 8, I-35131, Padova, Italy}
\affiliation{INFN, Sezione di Padova, via Marzolo 8, I-35131, Padova, Italy}
\affiliation{INAF- Osservatorio Astronomico di Padova, \\ Vicolo dell Osservatorio 5, I-35122 Padova, Italy}
\affiliation{Gran Sasso Science Institute, Viale F. Crispi 7, I-67100 L'Aquila, Italy}

\begin{abstract}

We extend the Schr\"odinger approach to large-scale structure formation
beyond the Newtonian regime by working at first post-Friedmann (1PF)
order. The standard Schr\"odinger--Poisson system gives a useful
reformulation of the dynamics of a self-gravitating pressureless fluid,
but it corresponds to the leading post-Friedmann, or Newtonian, limit.
It therefore misses the relativistic corrections that enter at
next-to-leading order and become relevant on horizon scales and for
high-precision cosmological surveys.

Starting from the 1PF continuity and Euler equations in a flat
$\Lambda$CDM background, we identify the conserved density variable
associated with covariant mass conservation. In terms of this variable,
the continuity equation takes a Newtonian-like conservative form.
However, even for vanishing covariant vorticity, the spatial velocity
field in the cosmological frame contains a transverse 1PF component.
Thus the full 1PF mass flux cannot be represented solely by the gradient
of a scalar phase.

We show that the Schr\"odinger-like formulation at 1PF order requires an
effective vector potential fixed by this transverse velocity component.
This vector potential contains the post-Friedmann metric vector
perturbation, related to relativistic frame-dragging effects, together
with nonlinear scalar terms required by the zero-vorticity condition.
Equivalently, when the equation is written in scalar form, these
corrections appear as an imaginary contribution to the effective
potential. At leading order, our system reduces to the usual
Schr\"odinger--Poisson formulation, while at 1PF order it provides a
relativistic extension of the Schr\"odinger description of cold matter
dynamics.
\end{abstract}

\maketitle


\section{Introduction}

A wide range of independent cosmological observations, including measurements of cosmic microwave background (CMB) anisotropies \cite{Planck}, large-scale clustering \cite{DESI_fullshape}, baryon acoustic oscillations (BAO) \cite{DESI_BAO}, and Type Ia supernovae \cite{Type_Ia_supernovae}, have converged toward a consistent description of the Universe, that the 
spatial curvature parameter $\Omega_k$ is tightly constrained by current observations to be consistent with zero at the level of $|\Omega_k| \lesssim \mathcal{O}(10^{-3})$ \cite{Planck, DESI_fullshape, DESI_BAO, Type_Ia_supernovae}, and whose energy budget, beyond photons, baryons, and neutrinos, is dominated by cold dark matter and a cosmological constant, this description of the Universe is commonly referred to as the concordance $\Lambda$CDM model \cite{concordance_model}. 
Within this framework, the large-scale structure (LSS) of the Universe emerges from the gravitational instability of primordial density fluctuations, which provide the initial conditions for structure formation. The large-scale structure of the Universe contains a wealth of cosmological information, as its statistical properties are sensitive to the initial conditions, the growth of structure, and the background expansion history, see, e.g., \cite{Peebles1980}. The seminal 1946 paper by Lifshitz \cite{Lifshitz:1945du}, pioneered the systematic study of the evolution of linear cosmological perturbations in an expanding background. It established the theoretical foundation for modern cosmological perturbation theory and subsequent studies of structure formation. The cosmological perturbation theory, see, e.g., \cite{PhysRevD.22.1882, Kodama:1984ziu, Wands2008}, provides a systematic description of linear and weakly nonlinear density fluctuations around a homogeneous and isotropic background when the amplitudes of the fluctuations are relatively small, but breaks down on highly nonlinear scales, under the assumption that cold dark matter (CDM) can be modeled as a self-gravitating, pressureless fluid, see, e.g., \cite{Blumenthal1984Natur, Peebles1982ApJ}. 

Throughout history, many analytical approaches to the formation of large-scale structures (LSS) have been developed based on cosmological perturbation theory, and these methods fall into two broad classes, namely the Eulerian and Lagrangian perturbation theory (for a review, see, e.g., \cite{Bernardeau_2002}).
Eulerian perturbation theory (EPT) perturbs macroscopic fluid quantities such as the density and velocity fields and examines their evolution at fixed spatial positions. Within the Eulerian framework, analytical approaches range from simple first-order perturbation theory (linearized fluid approach), e.g., \cite{Lifshitz:1945du, Peebles1980}, to higher-order approaches, e.g., \cite{Catelan1994, Sahni_1995}. First-order Eulerian perturbation theory is famous for being simple and robust when the density fluctuations are very small compared to the average matter density at early times or on large scales, but it may lead to unphysical negative matter densities at late times or on small scales. 
Lagrangian perturbation theory (LPT), e.g., \cite{Buchert1993, Buchert1994, Bouchet1995, Catelan:1994ze}, in which the particle displacement field is perturbed, is considered a competitive alternative to Eulerian perturbation theory. The Zel’dovich approximation (ZA) \cite{Zeldovich, Zelrev1989} provides an approximate treatment of the nonlinear dynamics of self-gravitating collisionless particles, which is a simple yet effective approach for capturing the main features of the clustering pattern, see \cite{colesTZ}, as long as there is no crossing of the particle trajectories. When the particle trajectories cross, the density will become locally infinite. This is called shell-crossing or caustic, and is a consequence of the fact that when two (or more) particles coming from different Lagrangian positions get to the same Eulerian position, the map between the Lagrangian position and the Eulerian position can no longer be one-to-one inverted.
After shell-crossing, the Zel’dovich approximation would imply that particles continue their motion along their straight trajectories and pass through the caustic, which makes the Zel’dovich approximation break down, because the gravitational intersection with the neighbouring particles is certainly going to modify their motion, and eventually they will be stabilized by gravity. To overcome this limitation, there is an extension to the Zel’dovich approximation by adding an artificial viscosity term in the Euler equation to mimic the gravitational forces between neighboring particles, dubbed ``adhesion approximation" \cite{Gurbatov:1989az}. Comparisons of the adhesion approximation with N-body simulations have shown good agreement even in the nonlinear regime, see \cite{1994ApJ...428...28M, Kofman:1991wt}.
A further significant improvement in describing the overall properties of the density and velocity fields over the Zel’dovich approximation is achieved by going to second-order Lagrangian perturbation theory (2LPT) for the fluid element paths, e.g., \cite{Buchert:1997dr, Melott:1994ah, Bouchet1995}, which includes the gravitational tidal effects as a correction to the Zel’dovich displacement. Compared to Eulerian perturbation theory, perturbations in the particle displacement field rather than the density and velocity fields capture more nonlinear information about the density and velocity fields (for more detailed discussions see, e.g., \cite{Bernardeau_2002}). 
Developed more recently, there are also other perturbation-theory-based formalisms such as the Effective Field Theory of Large-Scale Structure (EFTofLSS) \cite{Carrasco_2012, Carrasco:2013mua, Porto:2013qua}. EFTofLSS differs from standard perturbation theory by additional terms that encode relevant short-distance physics, such as the effective sound speed, viscosity, and stochastic pressure.  

In 1993, a wave-mechanical approach to the cosmic structure formation was suggested by Widrow and Kaiser \cite{Widrow:1993qq}, where the dynamics of a self-gravitating pressureless fluid can be described using a wave-mechanical formalism. Since then, it has attracted considerable interest and been extensively discussed and developed, with many extensions and applications; see, e.g. \cite{Coles:2002sj, Coles:2001fw, Short_2006a, Short_2006b, Gallagher_2022, Szapudi_2003, Johnston2010, coles2002wavemechanicslargescalestructure, coles2025, Uhlemann:2014npa, Uhlemann:2018gzz, Gough_2022, Uhlemann_Kopp_2014}. In this approach, collisionless CDM is modeled as a complex scalar field, with the dynamics governed by coupled Schrödinger and Poisson equations. This framework provides a useful approach for studying LSS formation driven by gravitational instability due to its advantage in smoothing the singularity of the density field at shell-crossing and in explaining the approximate log-normal form of the density fluctuations distribution function evolved from Gaussian initial conditions, as shown by Coles (2002)  \cite{Coles:2001fw} and Coles \& Spencer (2003) \cite{Coles:2002sj}.  The free-particle solution of the Schr\"odinger equation approach actually arrives at the trajectory found in the Zel’dovich approximation, and it is essentially an alternative to the adhesion model, in which the quantum pressure term acts as a viscosity term; see Short \& Coles (2006a,b) \cite{Short_2006a, Short_2006b}. Johnston, Lasenby \& Hobson (2010) \cite{Johnston2010} studied the analytic solution for the evolution of a compensated spherical overdensity. The evolution of isolated spherical cosmic voids has been studied by Gallagher \& Coles (2022) \cite{Gallagher_2022}, and following this work, Coles \& Gallagher (2025) \cite{coles2025} discussed the interpretation of the “quantum pressure” as a pressure in a true analogy with a classical fluid. Szapudi \& Kaiser (2003) \cite{Szapudi_2003}  introduced a theory of nonlinear cosmological perturbations with the Schr\"odinger equation, which in the limit of $\hbar\rightarrow0$, is a viable alternative to the Vlasov (or collisionless Boltzmann) equations that remain valid after shell-crossing. Uhlemann, Rampf, Gosenca \& Hahn (2019) \cite{Uhlemann:2018gzz} introduced a semiclassical method for evolving CDM by using the propagator, which encodes the transition amplitude of the wave function for the particle trajectories. The propagator follows from the Schr\"odinger equation using perturbation theory, where the leading-order propagator is the semiclassical equivalent of the Zel’dovich approximation, and for sufficiently small $\hbar$ the corresponding propagator solutions closely resemble LPT, with the addition that spurious vorticity is avoided and the dynamics at shell-crossing is regularised. 
This approach, although originally developed as a method to study particle-like CDM using wave mechanics, naturally describes the evolution of inherently wave-like dark matter (e.g. \cite{Brook_2022, PhysRevD.102.083518, Hui_2017}).  Such wave dark matter could be made of light bosonic dark matter with a long de Broglie wavelength, including, for example, the QCD axion \cite{PecceiQuinn, Abbott:1982af, Preskill:1982cy, Dine:1982ah}, which was introduced as a solution to the strong CP problem in quantum chromodynamics (QCD), or fuzzy dark matter (FDM) like some axion-like particles (ALP) (see, e.g., \cite{Hui2021,Kimball:2023vxk,Ferreira_2021}, and references therein).
 
At the same time, modern LSS surveys are designed to cover larger volumes and measure with higher statistical precision, enabling tighter constraints on cosmological parameters and a more detailed study of structure formation across cosmic time. For instance, Euclid probes structure formation over half of the age of the Universe; it will provide high-resolution optical imaging,  near-infrared imaging and spectroscopy, over about 14000 deg$^2$ of extragalactic sky, exploring the history of the expansion and the structure formation \cite{Euclid:2024yrr, EuclidTheoryWorkingGroup:2012gxx, Euclid:2014mgp}; DESI obtains more than 30 million galaxy and quasar redshifts to measure the BAO feature and determine the matter power spectrum, including redshift space distortions \cite{DESI:2016fyo}. Future surveys like Square Kilometre Array (SKA) \cite{Maartens:2015mra, SKA:2018ckk} will reach the scale of the order of the Hubble horizon with huge advances in resolution, sensitivity, and survey speed as a radio; the Large Synoptic Survey Telescope (LSST) \cite{LSSTDarkEnergyScience:2012kar} is designed to
image a substantial fraction of the sky in six optical bands to meet the requirements of a broad range of science goals in astronomy, astrophysics and cosmology, including the study of dark energy; the Nancy Grace Roman Space Telescope \cite{Roman} will have a field of view at least 100 times larger than Hubble’s and will look at billions of cosmic objects to explore how planets, stars, and galaxies form and develop over time. It is therefore essential that the theoretical framework used for predictions and data interpretation achieves a comparable level of accuracy. Going beyond the Newtonian approximation in the studies of LSS formation in order to take into account more GR effects can be crucial in this era of high-precision and Hubble horizon scale surveys. Many works have been done to investigate the impact of general relativistic nonlinear effects on cosmological structure and observables, see, e.g., \cite{48Chisari:2011iq, Green:2011wc, 50Bruni:2011ta, 52Bartolo2010, 53Bruni:2013qta, 54Bruni:2014xma, 55Kopp:2013tqa, 57Green:2014aga, 58Villa:2014foa, 59Rampf:2014xpt}. The effect of primordial and GR-induced non-Gaussianities on the second-order matter density perturbation in a $\Lambda$CDM cosmology can be interpreted as an effective non-zero $f_{NL}$, as shown by the relativistic second-order analyses in \cite{52Bartolo2010,53Bruni:2013qta}, cf. \cite{54Bruni:2014xma}. GR nonlinear effects give rise to a small gravitational slip and produce vector and tensor potentials from first-order scalar perturbations, which are indispensable for accurately modeling light propagation such as the redshift-space distortion, the gravitational lensing, and the Sachs-Wolfe effect, see, e.g., \cite{Yoo:2014sfa}. In \cite{Green:2010qy, Green:2011wc} a relatively straightforward “dictionary” mapping Newtonian dust cosmologies into general relativistic ones was proposed, and an ordering scheme to determine how the resulting metric and matter distribution fail to satisfy the Einstein field equations was introduced, thus providing a quantitative measure of the deviation between Newtonian and relativistic dust cosmologies (cf. \cite{48Chisari:2011iq}).
Various works have applied the traditional post-Newtonian (PN) expansion in cosmology. The PN equations of motion for particles in an expanding universe are derived, see \cite{70PhysRevLett.61.2175, 72, 73PhysRevD.53.681}. The Lagrangian approach to the PN analysis is given using the 3+1 (ADM) formalisms and the synchronous and comoving
gauge as detailed in \cite{75Takada1997APL} and \cite{76Matarrese_1996}. 
In \cite{Kofman_1995}, it is shown that the equations governing gravitational instability in an expanding universe are intrinsically non-local, implying that a consistent treatment requires retaining the non-local contributions of the Weyl tensor, which enter at the first post-Newtonian order. A closely related analysis is presented in \cite{Bertschinger:1994nc}, where the evolution equations for the electric and magnetic parts of the Weyl tensor for cold dust were derived, and it was found that the magnetic part does not vanish, implying that the Lagrangian evolution of the fluid is not purely local.
Notable works following the PN method of Szekeres et al. \cite{Szekeres_2000, Szekeres:2000ki} showed that standard Newtonian cosmological theory is inadequate, the Bianchi identities are not obtainable from the field equations, there is no well-posed initial value problem, and that the full Bianchi identities do not emerge in the linear approximation. The Bianchi identities, which express the covariant conservation of the Einstein tensor, ensure the consistency of the Einstein field equations and imply local energy--momentum conservation. Their absence in Newtonian gravity reflects the lack of an underlying covariant structure and underlies the limitations of the Newtonian approximation on horizon scales.
Based on this understanding, Milillo et al. \cite{Missinglink}, retained all scalar terms up to order $1/c^4$, and all vector terms up to $1/c^5$ in the equations, instead of peeling-off the different order, to present a set of nonlinear resummed equations up to the first order which retain in full the nonlinearity of Newtonian theory on small scales and all linear relativistic perturbation theory on large scales, which is called the nonlinear post-Friedmann (PF) framework. The PF approximation scheme is a generalization of the post-Minkowskian approximation to cosmology, while incorporating the
small velocity assumption of the PN approximation, thereby the approximate nonlinear general relativistic dynamics in $\Lambda$CDM cosmology are considered. The PF scheme provides a unified framework for describing the evolution of large-scale structures across all cosmologically relevant scales, assuming a flat $\Lambda$CDM background and a pressureless matter fluid. In this framework, the Newtonian cosmology equations naturally emerge as the leading-order, or zero post-Friedmann order (0PF), of the Einstein field equations. The next-to-leading order, or first post-Friedmann order (1PF), terms introduce general relativistic corrections, which include terms quadratic in the Newtonian variables, and proper linear GR terms. The linearization of the 1PF approximation recovers the first-order relativistic perturbation theory. However, beyond these linear terms, the 1PF scheme also contains second and higher order contributions within the standard perturbative expansion. Tensor modes at this order are purely nonlinear and non-dynamical; they do not correspond to propagating gravitational waves but instead represent distortions of the spatial slices. Close in spirit to \cite{Missinglink}, a mix between standard cosmological perturbation theory and PN approximation, \cite{Carbone:2004iv} studied the evolution of cosmological perturbations, using a hybrid approximation scheme which upgrades the weak-field limit of Einstein field equations to account for post-Newtonian scalar and vector metric perturbations and for leading-order source terms of gravitational waves, while including also the first and second-order perturbative approximations. A full-sky derivation of weak lensing observables in the PF formalism is done in \cite{Gressel_2019}, the cosmic structure formation on all scales from the perspective of GR is studied in \cite{Rampf_2016}, considering the PF approximation scheme, and the PF approach is applied to $f(R)$ gravity and a coherent derivation of the equations used in $f(R)$ N-body simulations is provided in \cite{Thomas_2015}.

The wave-mechanical approach, sometimes referred to as the Schr\"odinger approach to the structure formation, is a powerful method for studying the cosmic structure formation. However, the governing equations in this framework, the Schr\"odinger–Poisson system, are inherently Newtonian, which limits the analysis to non-relativistic dynamics. Nonetheless, the Schr\"odinger method can be generalized beyond the Newtonian regime (cf. \cite{Widrow:1996eq}). The objective of this paper is to extend the wave-
mechanical approach by incorporating some general relativistic corrections to the system, such that to the leading order it reduces to the standard Schr\"odinger-Poisson system and includes general relativistic corrections at the next-to-leading order. For this purpose, we will work within a nonlinear relativistic framework for the evolution of the LSS, called the post-Friedmann (PF) approximation. This scheme combines the weak-field approximation with the assumption of small peculiar velocities, working in a flat $\Lambda$CDM background and treating CDM as a self-gravitating pressureless fluid. In this framework, the leading, or zero post-Friedmann order (0PF) reproduces the Newtonian regime, while in the next-to-leading order, or first post-Friedmann order (1PF), general relativistic corrections are incorporated.

The paper is organized as follows. In Section II, we introduce the various metric terms and matter variables, and present the relevant components of the Einstein field equations, the energy–momentum conservation equations, and the vorticity tensor in terms of these variables. In Section III, we construct the system of equations governing the wave-mechanical approach to LSS formation generalized beyond the Newtonian regime.

\section{Post-Friedmann hydrodynamic equations}

As we already pointed out above, in this work, we adopt the post-Friedmann formalism, which generalizes the post-Minkowskian (weak-field) approximation to cosmology, while incorporating the key assumption of the post-Newtonian (PN) approximation that peculiar velocities remain small, and thus providing a nonlinear relativistic framework for structure formation, valid across both small and large scales, in a universe filled with a pressureless fluid and a cosmological constant $\Lambda$. In this section, we briefly review the post-Friedmann approximation developed in \cite{Missinglink} and present the main equations that will be used in the following sections.

\subsection{Metric and matter variables}
The aim is to work with the full Einstein field equations and obtain a self-consistent approximate system at each perturbative order. We consider a homogeneous, isotropic, and spatially flat FLRW background, on top of which inhomogeneous perturbations are introduced, with the expansion parameter formally defined as $1/c$. All scalar and tensor terms are expanded up to order $1/c^4$, and vector terms up to order $1/c^5$. See also \cite{Missinglink}, where the same resummation is considered.

The Newtonian approximation corresponds to the leading, 0PF order, while the first relativistic effects appear at the 1PF order. In the post-Friedmann framework, our aim is to determine these relativistic correction terms. We denote the spacetime coordinates by $x^\mu = (ct, x^i)$, where $t$ is the cosmic time and $x^i$ are comoving spatial coordinates.
By expanding the inhomogeneous perturbations of the components of the metric tensor in the line-element
\begin{equation}
\begin{aligned}
 & -c^{2} d\tau ^{2} =ds^{2} =g_{\mu \nu } dx^{\mu } dx^{\nu }\ ,
\end{aligned}
\end{equation}
where $\tau$ is the proper time, into the sum of terms of different orders of $1/c$, different components of the metric we are considering up to $O\left(1 / c^4\right)$ can be written as,
\begin{equation}\label{metric}
\begin{aligned}
&g_{00} =-1-\frac{1}{c^{2}} 2\phi _{G} -\frac{1}{c^{4}} 2\left( \phi _{G}^{2} +D_{P}\right)\ ,\\
&g_{0i} =-\frac{1}{c^{3}} aP_{i}\ ,\\
&g_{ij} =a^{2}\left[ 1-\frac{1}{c^{2}} 2\phi _{G} +\frac{1}{c^{4}} 2\left( \phi _{G}^{2} +D_{P}\right)\right] \delta _{ij} +\frac{1}{c^{4}} a^{2} h_{ij}\ ,
\end{aligned}
\end{equation}
where the Greek indices run over 0,1,2,3 and denote spacetime coordinates, while Latin indices run over 1,2,3 and denote purely spatial coordinates, and $\delta_{ij}$ is the Kronecker delta. $a(t)$ is the scale factor of the FLRW background, which satisfies the Friedmann equations for the flat FLRW background
\begin{equation}\label{Friedmann}
\begin{aligned}
\frac{1}{c^2} H^2 =\frac{1}{c^2} \frac{8 \pi G}{3} \bar{\rho}+\frac{\Lambda}{3}\ ,\\
\frac{1}{c^2}\left[\dot{H}+H^2\right]=-\frac{1}{c^2} \frac{4 \pi G}{3} \bar{\rho}+\frac{\Lambda}{3}\ .
\end{aligned}
\end{equation}
Here, we have defined  $\phi_G$, $D_P$ and $P_i$ as
\begin{equation}
\begin{aligned}
\phi_G=-\frac{1}{2}\left[U_N+V_N+\frac{2}{c^2}(U_P+V_P)\right]\ ,\\
\frac{1}{c^2}D_P=-\frac{1}{2}\left[U_N-V_N+\frac{2}{c^2}(U_P-V_P)\right]\ ,
\end{aligned}
\end{equation} and
\begin{equation}
P_i=B^N_i+\frac{1}{c^2}B^P_i\ ,
\end{equation}
that is, they are resummed variables, where $U$ and $V$ are the gravitational potentials, and $B_i$ is the vector potential, indices N and P label 0PF and 1PF quantities respectively, for details see \cite{Missinglink}. In particular, $\phi_G$ is a generalized gravitational potential, and $D_P$ defines a nonlinear PF quantity, representing the gravitational slip, which is negligible in the Newtonian and in the linear regimes.
We work in the Poisson (or conformal Newtonian) gauge, e.g., \cite{Ma:1995ey, Bardeen1980}, where $P_i$ is divergenceless vectorial potential, satisfying $P_i^{,i} = 0$, and $h_{ij}$ is transverse and trace-free, satisfying $h^{i}{}_{i} = 0$ and $h_{ij}^{,i} = 0$, representing pure tensor modes. Therefore, there remain six degrees of freedom at each order. However, the leading order Einstein field equations impose $U_N=V_N$, and the tensor modes arise only from 1PF order. 

The metric can be viewed as a cosmological generalization of Chandrasekhar’s post-Newtonian hydrodynamic metric \cite{1965ApJ...142.1488C}. Having defined the metric variables, we turn to the matter sector.
Dimensionless four-velocity is defined as $u^\mu:=d x^\mu/c d \tau$, which satisfies the normalization condition $g_{\mu \nu} u^\mu u^\nu=-1$. The physical position of a fluid element is $\mathbf{r}=a \mathbf{x}$, where $\mathbf{x}$ represents the background comoving coordinates. Therefore, we define the peculiar physical velocity as $v^{i} =adx^{i}/{dt}$, we also define $v_{i} :=\delta _{ij} v^{j}$ and $\delta _{ij}v^jv^i=v^2$. Therefore, the four-velocity can be written as
\begin{equation}
u^i
=\frac{d x^i}{c d t} \frac{d t}{d \tau}=\frac{v^i}{c a} u^0\;.
\end{equation}
Using the four-velocity normalization condition and keeping terms up to the order of $1/c^{4}$, the components of four-velocity are
\begin{equation}\label{four-velocity}
\begin{aligned}
 & u^{i} =\frac{1}{c}\frac{v^{i}}{a} +\frac{1}{c^{3}}\frac{v^{i}}{a}\left(-\phi _{G} +\frac{1}{2} v^{2}\right)\ , \\
 & u^{0} =1+\frac{1}{c^{2}}\left(-\phi _{G} +\frac{1}{2} v^{2}\right) +\frac{1}{c^{4}}\left[\frac{1}{2}\phi _{G}^{2} -D_{P}-\frac{5}{2} v^{2}\phi _{G} +\frac{3}{8} v^{4} -P_{i}v^{i}\right]\ ,\\
 & u_{i} =\frac{av_{i}}{c} +\frac{a}{c^{3}}\left[ -P_{i}  -3v_{i} \phi _{G} +\frac{1}{2} v_{i} v^{2}\right]\ ,\\
 & u_{0} =-1+\frac{1}{c^{2}}\left(-\phi _{G} -\frac{1}{2} v^{2}\right) +\frac{1}{c^{4}}\left[-D_{P} -\frac{1}{2} \phi _{G}^{2} +\frac{3}{2} v^{2} \phi _{G}  -\frac{3}{8} v^{4}\right]\ .
\end{aligned}
\end{equation}
Considering the Einstein field equations including the cosmological constant $\Lambda$,
\begin{equation}
G_\nu^\mu=R^{\mu}_{\nu}-\frac{1}{2}R\delta^{\mu}_{\nu}=\frac{8 \pi G}{c^4} T_\nu^\mu-\Lambda \delta_\nu^\mu \ ,
\end{equation}
where $R^{\mu}_{\nu}$ is the Ricci tensor and $R$ is the Ricci scalar. Note that the stress energy tensor $T_\nu^\mu$ describes a single pressureless (dust) fluid, i.e., the Cold Dark Matter (CDM) component. The energy–momentum tensor is 
\begin{equation}
T_\nu^\mu=c^2\rho u^\mu u_\nu\ ,
\end{equation}
where $\rho$ is the CDM density. The density contrast is defined as
\begin{equation}
\delta:=\frac{\rho-\bar\rho}{\bar\rho}\ ,
\end{equation}
with $\bar\rho$ representing the background matter density.
Expanding the Einstein field equations in powers of $1/c$ with the metric given in Eq.\ (\ref{metric}), we obtain the different components of the Einstein field equations \cite{Missinglink},
\begin{align}
G_0^0+\Lambda=\frac{8 \pi G}{c^4} T_0^0 \rightarrow & \frac{1}{c^2} \nabla^2 \phi_G-\frac{1}{c^4}\left[\nabla^2 D_P+3 a^2\left(\frac{\dot{a}}{a} \dot{\phi}_G+\left(\frac{\dot{a}}{a}\right)^2 \phi_G\right)-\nabla^2 \phi_G^2+\frac{5}{2} \phi_{G,i} \phi_G^{,i}\right] \notag\\
& =\frac{1}{c^2} 4 \pi G a^2 \bar{\rho} \delta+\frac{1}{c^4} 4 \pi G a^2 \bar{\rho}(1+\delta) v^{ 2}\ , 
\end{align}
\begin{align}
{\left[G_i^j+\Lambda \delta_i^j=\frac{8 \pi G}{c^4} T_i^j\right]_{\text {trace }} \rightarrow } & \frac{1}{c^4}\left[4 \nabla^2 D_P+\phi_{G, k} \phi_G^{, k}+6 a^2\left(4 \frac{\dot{a}}{a} \dot{\phi}_G+2 \frac{\ddot{a}}{a} \phi_G+\left(\frac{\dot{a}}{a}\right)^2 \phi_G+\ddot{\phi}_G\right)\right] \notag \\
& =\frac{1}{c^4} 8 \pi G a^2 \bar{\rho}(1+\delta) v^{ 2}\ ,
\end{align}
\begin{align}
{\left[G_i^j+\Lambda \delta_i^j=\frac{8 \pi G}{c^4} T_i^j\right]_{\text {tracefree }} \rightarrow } & \frac{1}{c^4}\left\{2\left(D_{P, i}^{, j}-\frac{1}{3} \nabla^2 D_P \delta_i^j\right)+2 \phi_{G, i} \phi_G^{, j}-\frac{2}{3} \phi_{G, k} \phi_G^{, k} \delta_i^j\right. \notag\\
& \left.\ \ \ \ \ -a\left[\frac{\dot{a}}{a}\left(P_i^{, j}+P_{, i}^j\right)+\frac{1}{2}\left(\dot{P}_i^{, j}+\dot{P}_{, i}^j\right)\right]+\frac{1}{2} \nabla^2 h_i^j\right\} \notag \\
& =-\frac{1}{c^4} 8 \pi G a^2 \bar{\rho}(1+\delta)\left(v_i v^{j}-\frac{1}{3} v^{ 2} \delta_i^j\right)\ ,
\end{align}
\begin{align}
G_i^0=\frac{8 \pi G}{c^4} T_i^0 \rightarrow & \frac{1}{c^3}\left[\frac{1}{2 a} \nabla^2 P_i+2 \frac{\dot{a}}{a} \phi_{G, i}+2 \dot{\phi}_{G, i}\right]-\frac{1}{c^5}\left\{-2 \frac{\dot{a}}{a} D_{P, i}+2 \dot{D}_{P, i}\right. \notag \\
& \left.+\frac{1}{2 a}\left[4 a\left(\dot{\phi}_G \phi_{G, i}+2 \phi_G \dot{\phi}_{G, i}\right)+8 \dot{a} \phi_G \phi_{G, i}+2 P_{k, i} \phi_{G, k}-P_i \nabla^2 \phi_G-2 P_k \phi_{G, k i}\right]\right\} \notag \\
& =-\frac{1}{c^3} 8 \pi G a \bar{\rho}(1+\delta) v_i-\frac{1}{c^5} 8 \pi G a \bar{\rho}(1+\delta)\left[v_i\left(v^{2}-4 \phi_G\right)-P_i\right]\ .
\end{align}
The dot denotes partial differentiation with respect to coordinate time $t$ and the comma denotes partial differentiation with respect to the background comoving coordinates $x^i$.
In all equations, we retain the first two terms in the expansion in powers of $1/c$, and to obtain equations governing the inhomogeneous quantities, we have subtracted the background part, i.e., Eq.\ (\ref{Friedmann}), from the full set of equations. The above equations are only affected by the cosmological constant through their coupling with the Hubble expansion, since the background dynamics have been subtracted.
Retaining the leading order terms in the $1/c$ expansion, i.e., the 0PF order, the Einstein field equations reduce to the equations of Newtonian cosmology on small scales. Conversely, upon linearising the system, corresponding to the 1PF approximation in the linear regime, one recovers standard first-order relativistic perturbation theory on large scales.

\subsection{1PF continuity equation and Euler equation}

Consistency of Einstein field equations with the contracted Bianchi identities enforces the covariant conservation of the energy–momentum tensor, i.e,
\begin{equation}
T^{\nu}{ }_{\mu;\nu}=0,
\end{equation}
where the time component gives the continuity equation, while the space component gives the momentum conservation equation.
From the Bianchi identity, considering the time component and keeping all terms up to $1/c^2$ order, we obtain the continuity equation to the first post-Friedmann order (1PF)
\begin{equation}\label{continuity1}
\frac{d \delta}{d t}+\frac{v_{, i}^i}{a}(\delta+1)-\frac{1}{c^2}\left[(\delta+1)\left(\frac{1}{a} v^j \phi_{G,j}+\frac{\dot{a}}{a} v^2+3 \frac{d \phi_G}{d t}\right)\right]=0\ ,
\end{equation}
and the space component gives the Euler equation to the 1PF order
\begin{equation}\label{Euler1}
\begin{aligned}
& \frac{d v_i}{d t}+\frac{\dot{a}}{a} v_i+\frac{\phi_{G, i}}{a}+\frac{1}{c^2}\left[-3v_i \frac{d\phi_G}{d t}+\frac{1}{a} \phi_{G, i}\left(4\phi_G+v^2\right)+\frac{1}{a} D_{P, i}-\frac{1}{a} v_i v^j \phi_{G, j}-\frac{\dot{a}}{a} v^2 v_i-\frac{1}{a} \frac{d}{d t}\left(a P_i\right)+\frac{P_{j, i}v^j}{a}\right]=0 ,
\end{aligned}
\end{equation}
where the convective derivative for an arbitrary quantity $Q$ is defined as
\begin{equation}
\frac{d Q}{d t}\equiv\dot{Q}+\frac{v^i Q_{,i}}{a}.
\end{equation}
Note that we have subtracted the background continuity equation $\dot{\bar{\rho}}=-3 H \bar{\rho}$ from the above equations, using the Newtonian parts to simplify the 1PF order contributions. Here, Eqs.\ (\ref{continuity1}) and (\ref{Euler1}) are the hydrodynamic equations of motion of a pressureless fluid in the post-Friedmann approximation up to 1PF order.

\subsection{Vorticity conservation}

In GR, the vorticity tensor is given by (see, e.g., \cite{Ellis:1971pg, Misner:1973prb})
\begin{equation}
\begin{aligned}
\omega _{\mu \nu } =\frac{1}{2}\left( h_{\mu }^{\gamma } h_{\nu }^{\sigma } -h_{\nu }^{\gamma } h_{\mu }^{\sigma }\right) u_{\gamma ;\sigma },
\end{aligned}
\end{equation}
with the projection tensor, $\displaystyle h_{\mu }^{\gamma } =\delta _{\mu }^{\gamma } +u_{\mu } u^{\gamma }$, into rest 3D space of an observer moving with four-velocity $u^\mu$. The vorticity characterizes the local rotational motion of fluid elements relative to the local rest frame defined by the fluid four-velocity.

Under the post-Friedmann approximation, using the four-velocity, which we have found 
in Eq.\ (\ref{four-velocity}) and keeping terms up to the 1PF order, the non-zero components of the vorticity tensor expanded up to 1PF order are the spatial components,
\begin{equation}
\begin{aligned}
2\omega _{ij} & =u_{i;j} -u_{j;i} +u_{i} u^{\gamma } u_{\gamma ;j} -u_{j} u^{\gamma } u_{\gamma ;i}\\
& =\frac{1}{c} a(v_{i,j} -v_{j,i} )\\
& \ \ \ \ \ +\frac{1}{c^{3}} a\left[\left(\frac{1}{2} v^{2} -3\phi _{G}\right) (v_{i,j} -v_{j,i} )+v_{i}\left(\frac{1}{2} v^{2} -3\phi _{G}\right)_{,j} -v_{j}\left(\frac{1}{2} v^{2} -3\phi _{G}\right)_{,i} -P_{i,j} +P_{j,i}\right]
\end{aligned}
\end{equation}
and the mixed time–space components of the vorticity tensor,
\begin{equation}
\begin{aligned}
2\omega _{0i} & =\frac{1}{c^{2}} v^{j}\bigl( v_{i,j} -v_{j,i}\bigr)+\frac{1}{c^{4}}\left[\left(\frac{1}{2} v^{2} -3\phi _{G}\right)_{,j} v_{i} v^{j} -\left(\frac{1}{2} v^{2} -3\phi _{G}\right)_{,i} v^{2} -v^{j}( P_{i,j} -P_{j,i})\right]\,. 
\end{aligned}
\end{equation}

It is well known that, see, e.g., \cite{Peebles1980, Bernardeau_2002}, during the structure formation, considering a pressureless perfect fluid, the velocity field remains vorticity-free, i.e., $\omega ^{\mu \nu } =0$, in the absence of shell-crossing, provided the initial conditions are vorticity-free. Since vector perturbations decay kinematically as the Universe expands, during inflation, any pre-existing vector modes are exponentially diluted by the rapid expansion, and quantum fluctuations of vector type are not amplified in the absence of additional vector fields, see, e.g., \cite{Mukhanov:2005sc}. As a consequence, there is no ``natural" mechanism for generating primordial vector perturbations in a single-field slow-roll inflationary model.
Given the vorticity-free initial condition, the conservation of the zero vorticity is guaranteed by the (Kelvin–Helmholtz) vorticity conservation laws (see, e.g., \cite{Ellisbook, Ehlers:1961xww}). The theorem states that, for an ideal barotropic fluid subject only to conservative forces, the circulation along a comoving closed loop is conserved. As a consequence, for a barotropic perfect fluid, vorticity cannot be generated dynamically if it is initially absent, i.e., vorticity can be generated only when the fluid is either non-barotropic or imperfect, in which cases the four-acceleration $u^\nu\nabla_\nu u_\mu$ does not vanish and acts as a source term for the propagation of the vorticity. 
In the nonlinear post-Friedmann approximation, the conservation of zero vorticity continues to hold for a pressureless dust fluid, as it follows from fully nonlinear general relativity as shown in \cite{Ellis:1971pg, Ellisbook, Ehlers:1961xww}.
Mind that in the linear perturbation theory, a zero vorticity velocity field usually indicates that the velocity field is curl-free, i.e., $\displaystyle v^{i} =\theta ^{,i}$. 
However, as we will see later in this section, this equivalence does not always persist beyond linear order. In nonlinear GR perturbation theory, vanishing vorticity does not necessarily imply that the three-dimensional spatial velocity field is curl-free.

Since the absence of vorticity is a consistent solution prior to shell-crossing, we apply the zero vorticity condition to our velocity field, which leads to a velocity field that can be decomposed to the 1PF order as
\begin{equation}
v_{i} =\theta _{,i} +v_{\bot i}\ ,
\end{equation}
with
\begin{equation}
\begin{aligned}
v_{\bot i} & =\frac{1}{c^{2}}\left\{P_{i}-v_{i}\left(-3\phi_{G} +\frac{1}{2} v^{2}\right) + \nabla ^{-2}\left[ v_{j}\left(-3\phi_{G} +\frac{1}{2} v^{2}\right)\right]_{,i}^{,j}\right\}\ .
\end{aligned}
\end{equation}
Consequently, considering that the vorticity vanishes, the transverse (divergence-free) component of the velocity field appears only at $\mathcal{O}\left( 1/c^{2}\right)$ in the post-Friedmann expansion. This component is therefore a genuinely 1PF order variable, satisfying $\nabla ^{i} v_{\bot i}=0$. Thus, to the 1PF order, a velocity field with zero vorticity can be written as
\begin{equation}
v_{i} =\theta _{,i} +\frac{1}{c^{2}}\vartheta_i\ ,
\end{equation}
defining
\begin{equation}\label{theta}
\frac{1}{c^2}\vartheta_i=\frac{1}{c^{2}}\left\{P_{i} -\theta _{,i}\left( -3\phi_{G} +\frac{1}{2} \theta _{,k} \theta ^{,k}\right) + \nabla  ^{-2}\left[ \theta _{,j}\left(-3\phi_{G} +\frac{1}{2} \theta _{,k} \theta ^{,k}\right)\right]_{,i}^{,j}\right\}\ .
\end{equation}
 Note that at 0PF order, a velocity field with vanishing vorticity reduces to the gradient of a scalar, which is called the velocity potential. In particular, in the Newtonian regime, zero vorticity implies a curl-free velocity field. However, at higher orders in the nonlinear post-Friedmann expansion, the velocity field acquires a transverse vector component even under the assumption of vanishing vorticity. This is due to the fact that the spatial curl of the peculiar velocity field depends on the chosen reference frame (or observer), whereas the vorticity tensor $\omega_{\mu \nu}$ is a covariant quantity defined directly from the four-velocity field.

Taking the curl of Eq.~(\ref{theta}), we are able to recast the curl part of the velocity field up to 1PF order into
\begin{equation}\label{curl_theta}
\frac{1}{c^{2}} \vartheta _{i,k} \varepsilon ^{ijk} =\frac{1}{c^{2}}\left[ P_{i,k} \varepsilon ^{ijk} -\frac{1}{a^{3}} \theta _{,i}\left(\sqrt{-g} u^{0}\right)_{,k} \varepsilon ^{ijk}\right]\ ,
\end{equation}
where we used
\begin{equation}
\sqrt{-g} u^{0} =a^{3}\left[1-\frac{1}{c^{2}}\left( 3\phi _{G} -\frac{1}{2} v^{2}\right)\right]\ .
\end{equation}

The first term on the RHS of Eq.~(\ref{curl_theta}) is precisely the curl of the off-diagonal spacetime metric components $g_{0i}$, whose rotational structure encodes the frame-dragging effect through the mixing of temporal and spatial directions.
In curved spacetime, the motion of gravitating sources can induce what is known as frame-dragging \cite{Lense:1918zz}, a phenomenon where the spacetime geometry itself imparts angular momentum structure to the inertial frames. To understand this effect, consider an irrotational freely falling observer, along the observer’s worldline, one can introduce a local inertial frame in which the metric reduces to the Minkowski form and the Christoffel symbols vanish at the reference point, which indicates that the spatial components of the vorticity tensor are simply 
\begin{equation}
\omega_{ij}=u^{LIF}_{i,j}-u^{LIF}_{j,i}=\frac{1}{c}(v^{LIF}_{i,j}-v^{LIF}_{j,i})=0\ ,
\end{equation} 
thus, the vanishing of vorticity. However, in the presence of frame-dragging, the local tetrad, the set of axes defining the observer's inertial frame, is gradually twisted by the gravitational field as it is parallel transported through the curved spacetime. This twisting is not due to any local vorticity of the observer congruence, but rather to the cumulative effect of spacetime curvature along the trajectory, which is precisely captured by the nonlinear sector of the post-Friedmann expansion (cf. \cite{Thomas:2015kua, Frame_draggingBruni:2013mua, Beordo:2025cpw}). The result is striking: when compared to distant comoving observers (who define the cosmological background frame), the local observer appears to rotate, even though the motion is locally irrotational, i.e., when we go to the background comoving frame, it will result in a non-zero curl of the velocity field.
This notion is analogous in spirit to that of zero angular momentum observers (ZAMO)  \cite{Bardeen1970_ZAMO} in stationary, axisymmetric, asymptotically flat spacetimes. In both cases, an observer who is ``non-rotating" in a local sense can nonetheless rotate relative to a global reference frame due to the dragging of inertial frames by spacetime curvature. 
The frame-dragging effect may be related to the Weyl tensor; further discussion is provided in the Appendix.

The second term on the RHS of Eq.~(\ref{curl_theta}) is the cross product between the gradient of the velocity potential $\theta_{,i}$ and the gradient of $\sqrt{-g} u^{0}$, which represents the ``proper spatial volume element factor". This can be seen by considering an invariant volume element for a four‑dimensional spacetime, i.e.,
\begin{equation}
dV_{4D}=\sqrt{-g}d^4x\ ,
\end{equation}
which can be factorised as
\begin{equation}
dV_{4D}=\left(\sqrt{-g}u^0d^3x\right)d\tau\ .
\end{equation}
Therefore, the LHS is the coordinate invariant 4-volume element, and on the RHS, $d\tau$ is the observer’s proper time interval. Hence $\sqrt{-g}u^0d^3x$ is the proper spatial volume element, i.e., the factor $\sqrt{-g}u^0$ is the conversion factor from coordinate volume $d^3x$ to the proper spatial volume element,
\begin{equation}
dV_{proper}=\sqrt{-g}u^0d^3x\ .
\end{equation}
The second term on the RHS of Eq.~(\ref{curl_theta}) represents the misalignment between the direction of the velocity and that of the spatial gradient of the spatial volume element measured by a local observer. This can formally induce a curl to the velocity field as there are changes in relative displacement between neighboring particle elements due to the spatial inhomogeneity of the space volume. To understand this effect better, we can consider a spacetime element along the worldline of a free-falling observer with $u^{\mu}=u^0(1,v^i/ac)$. When the direction of the velocity $v^i$ is misaligned with the gradient of the proper spatial volume element, a physical asymmetry in the proper volume develops across the observer's local frame. Consequently, the local observer measures a velocity differential between either side of the worldline. To counteract this rotational structure and maintain a zero vorticity flow, the velocity field $v^i$ formally develops a curl.

\subsection{A coupled system of Bernoulli and continuity equations }
 
It is convenient to use the linear growth-factor $D(t)$, which is the growing mode solution of 
\begin{equation}
\ddot{D} +2H\dot{D} -4\pi G\overline{\rho } D=0\ ,
\end{equation}
which follows from linear perturbation theory in an expanding universe, as the evolution variable. Using $D(t)$ as the evolution variable simplifies the equations of motion and provides a unified description of the linear growth across different cosmological models, since $D(t)$ incorporates the effects of the background expansion history within linear perturbation theory.
Following \cite{Short_2006a, Short_2006b}, we define the comoving 
velocity field as
\begin{equation}
\begin{gathered}
V^{i}( x^{i} ,D) :=\frac{d x^{i}}{d D}=\frac{v^{i}}{a\dot{D}}\ ,
\end{gathered}
\end{equation}
where we also have defined
\begin{equation}
\begin{gathered}
V_{i}=\delta_{ij}V^j  \quad \quad {\rm and} \quad \quad V^2=\delta_{ij}V^jV^{i}\ \nonumber.
\end{gathered}
\end{equation}
Then we have
\begin{equation}
\dot{v}_{i} =\frac{\partial }{\partial t} (a\dot{D} V_{i} )=(\dot{a}\dot{D} +a\ddot{D} )V_{i} +a\dot{D}^{2}\frac{\partial V_{i}}{\partial D}\ .
\end{equation}
Using  the (linear) growth rate 
\begin{equation}\label{f}
f:=\frac{d\ln{D}}{d\ln{a}}\;,
\end{equation}
we have
$\dot{D}\equiv fHD\ $,
we can rewrite the continuity equation Eq.\ (\ref{continuity1}) and the Euler equation Eq.\ (\ref{Euler1}) in terms of the linear growth factor $D(t)$, i.e.,
\begin{equation}\label{continuity3}
\frac{\partial \delta }{\partial D} +V^{i} \delta _{,i} +(\delta +1)\left[ V_{,i}^{i} -\frac{1}{c^{2}}\left( 3\frac{\partial \phi _{G}}{\partial D} +4V^{j} \phi _{G,j} +a^2H^2 fDV^{2}\right)\right] =0\ ,
\end{equation}
and
\begin{equation}\label{Euler3}
\begin{aligned}
\frac{\partial V_{i}}{\partial D} +V^{j} V_{i,j} & =-\left[\frac{4\pi G\overline{\rho }}{H^{2} f^{2} D} -\frac{1}{c^{2}}\left(a^2H^2fDV^{2} +3\frac{\partial \phi _{G}}{\partial D} +4V^{j} \phi _{G,j}\right)\right] V_{i}\\
& \ \ \ \ -\left[\frac{1}{a^2H^2 f^{2} D^{2}} +\frac{1}{c^{2}}\left(\frac{4}{a^2H^2 f^{2} D^{2}} \phi _{G} +V^{2}\right)\right]\left( \phi _{G,i} +\frac{1}{c^{2}} D_{P,i}\right)\\
& \ \ \ \ -\frac{1}{c^{2}}\frac{1}{aHfD}\left(\frac{\partial P_{i}}{\partial D} +V^{j}( P_{i,j} -P_{j,i}) +\frac{1}{f D} P_{i}\right) \ .
\end{aligned}
\end{equation}
This is useful because, from now on, we are able to work directly in terms of a dimensionless evolution variable.
Retaining the 0PF order, the equation reduces to the Euler equation
\begin{equation}
\frac{\partial V_{i}}{\partial D} +V^{j} V_{i,j} =-\frac{4\pi G\overline{\rho }}{H^{2} f^{2} D} V_{i}-\frac{1}{a^2H^2 f^{2} D^{2}} \phi _{G,i}\ .
\end{equation}
The contributions from the vector potential is suppressed at 0PF order since it enters only at higher order in the post-Friedmann approximation, therefore do not contribute to the Newtonian matter flow.
However, note that even at 0PF order, the vector potential does not necessarily vanish for a perfect irrotational fluid when nonlinear effects are taken into account (cf. \cite{Thomas:2015kua, Frame_draggingBruni:2013mua, Beordo:2025cpw}), as it can be sourced by nonlinear coupling between density and velocity perturbations, i.e., at 0PF order, the fully nonlinear time-space components of the Einstein field equations indicate
\begin{equation}
\frac{1}{c^3} \nabla \times \nabla^2 P_i=-\frac{16 \pi G a^2}{c^3} \nabla \times\left(\rho v_i\right)\ .
\end{equation}
At 1PF order, corrections to the motion of the fluid particles beyond Newtonian dynamics are included, in particular, the effect of gravitational slip enters through $D_P$, and contributions from the vector potential $P_i$ are also included.

Applying the decomposition of the vorticity-free velocity field, $v_{i} =\theta _{,i} +\vartheta_{i} /c^2$, discussed in Section II, and introducing the rescaled variables
\begin{equation}
\begin{gathered}
\Theta _{,i}  =\frac{1}{a\dot{D}} \theta _{,i}\ ,\\ 
\frac{1}{c^{2}} \varTheta _{i} =\frac{1}{c^{2}}\frac{1}{a\dot{D}} \vartheta _{i} \ ,
\end{gathered}
\end{equation}
the comoving velocity field can be written as
\begin{gather}\label{decomposed V_i}
V_i=\Theta _{,i}+\frac{1}{c^{2}} \varTheta _{i}\ ,
\end{gather}
where $ \varTheta _{i}$ is then given by Eq.\ (\ref{theta}) as
\begin{equation}
\frac{1}{c^{2}} \varTheta _{i} =\frac{1}{c^{2}}\left\{\frac{1}{aH fD} P_{i} -\Theta _{,i}\left( -3\phi _{G} +\frac{1}{2}a^2H^2 f^{2} D^{2} \Theta _{,k} \Theta ^{,k}\right) +\nabla ^{-2}\left[ \Theta _{,j}\left( -3\phi _{G} +\frac{1}{2}a^2H^2 f^{2} D^{2} \Theta _{,k} \Theta ^{,k}\right)\right]_{,i}^{,j}\right\}\ .
\end{equation}
Substituting Eq.\ (\ref{decomposed V_i}) into Eq.\ (\ref{continuity3}) and Eq.\ (\ref{Euler3}), we obtain
\begin{equation}\label{continuity2}
\frac{\partial \delta }{\partial D} +\Theta ^{,i} \delta _{,i} +\nabla ^{2} \Theta (\delta +1)-\frac{1}{c^{2}}\left[ (\delta +1)\left( 3\frac{\partial \phi _{G}}{\partial D} +4\Theta ^{,k} \phi _{G,k} +a^2H^2fD \Theta ^{,k} \Theta _{,k}\right) -\varTheta^{i} \delta _{,i}\right] =0\ ,
\end{equation}
and
\begin{equation}\label{Euler4}
\begin{aligned}
 & \frac{\partial \Theta _{,i}}{\partial D} +\Theta ^{,j} \Theta _{,ij} +\frac{4\pi G\overline{\rho }}{H^{2} f^{2} D} \Theta _{,i} +\frac{1}{a^2H^2 f^{2} D^{2}} \phi _{G,i}\\
 & +\frac{1}{c^{2}}\left[\frac{1}{a^2H^2 f^{2} D^{2}} D_{P,i} -\left( 3\frac{\partial \phi _{G}}{\partial D} +4\Theta ^{,j} \phi _{G,j} +a^2H^2 fD\Theta _{,j} \Theta ^{,j}\right) \Theta _{,i} +\left(\frac{4}{a^2H^2 f^{2} D^{2}} \phi _{G} +\Theta _{,j} \Theta ^{,j}\right) \phi _{G,i} \right. \\
 & \ \ \ \ \ \ \ \ \left. +\frac{\partial \varTheta _{i}}{\partial D} +\Theta ^{,j} \varTheta _{i,j} +\varTheta ^{j} \Theta _{,ij} +\frac{4\pi G\overline{\rho }}{H^{2} f^{2} D} \varTheta _{i} -\frac{1}{aH fD}\left(\frac{1}{f D} P_{i} +\frac{\partial P_{i}}{\partial D} +\Theta ^{,j}( P_{i,j} -P_{j,i})\right)\right] =0\ .
\end{aligned}
\end{equation}
By taking the divergence and the inverse Laplacian, we are able to recast Eq.\ (\ref{Euler4}) into a Bernoulli-like form
\begin{equation}\label{Bernoulli}
\frac{\partial \Theta }{\partial D} +\frac{1}{2} \Theta ^{,j} \Theta _{,j}+\mathcal{V}=0
\end{equation}
where $\mathcal{V}$ is an effective potential expanded up to 1PF order
\begin{equation}
\mathcal{V}=\frac{4\pi G\overline{\rho }}{H^{2} f^{2} D}\Theta +\frac{1}{a^2H^2f^2D^2} \phi _{G} +\frac{1}{c^{2}}\mathcal{A}\ .
\end{equation}
At leading order, $\mathcal{V}$ reduces to the effective potential appearing in the Bernoulli-like equation in the Newtonian regime, see, e.g., Eq. (18) in \cite{Short_2006a}, while at next-to-leading order, i.e., at 1PF order, the post-Friedmann correction turns out
\begin{equation}
\begin{aligned}
\frac{1}{c^{2}}\mathcal{A} & =\frac{1}{c^{2}}\left\{\nabla ^{-2} \partial ^{i}\left[a^2H^2 f^{2} D^{2} \Theta ^{,j} \Theta ^{,k}( \Theta _{,j} \Theta _{,ki} -\Theta _{,i} \Theta _{,kj}) -\left( 3\frac{\partial \phi _{G}}{\partial D} +\Theta ^{,j} \phi _{G,j} +a^2H^2 fD\Theta _{,j} \Theta ^{,j}\right) \Theta _{,i}\right. \right. \\
 & \left. \left. \ \ \ \ \ \ \ \ \ \ \ \ \ \ \ \ \ \ \ +\left(\frac{4}{a^2H^2 f^{2} D^{2}} \phi _{G} -2\Theta _{,j} \Theta ^{,j}\right) \phi _{G,i}\right] +\frac{1}{a^2H^2 f^{2} D^{2}} D_{P} +{ \varTheta ^{i} \Theta _{,i}}\right\} \ ,
\end{aligned}
\end{equation}
which represents the 1PF correction to the effective potential. At linear order, it includes the contribution from $D_P$, which depends on the scalar potential associated with anisotropic stress (e.g., from relativistic species such as neutrinos in a $\Lambda$CDM framework). 
Beyond linear order, $\mathcal{A}/c^2$ encodes nonlinear post-Friedmann contributions that are absent in standard first-order cosmological perturbation theory, including the contributions form self-advection terms of the velocity perturbations such as $\varTheta ^{i} \Theta _{,i}$, nonlinear self-couplings of the metric perturbations such as $\phi _{G}\phi _{G,i}$, as well as mixed couplings between metric and velocity perturbations. These nonlinear terms are further modulated by the background expansion through the Hubble parameter $H$ and the linear growth rate $f$. 
We also note that Eq.\ (\ref{Bernoulli}) together with the continuity equation Eq.\ (\ref{continuity2}) give a coupled system of Bernoulli and continuity equations describing the dynamics of a pressureless fluid in the post-Friedmann approximation up to 1PF order, which is one of the main results of this work. This post-Friedmann correction $\mathcal{A}/c^2$ to the effective potential of the Bernoulli-like equation constitutes one of the main results of this work.

\section{The Schr\"odinger-like equation using PF approach }

Through the Madelung transformation \cite{Madelung:1927ksh}, the Schr\"odinger equation for a spinless non-relativistic particle can be reformulated as a system of two coupled partial differential equations for the amplitude and the phase of the wave function in polar form: a continuity equation for the probability density (the square of the amplitude) and an Euler-type equation for a velocity field defined from the phase, with an additional quantum potential term.
From a cosmological point of view, as pointed out by Kaiser \& Widrow \cite{Widrow:1993qq}, see also, e.g., \cite{Coles:2002sj, Coles:2001fw, Short_2006a, Short_2006b, Gallagher_2022, Szapudi_2003, Johnston2010, coles2002wavemechanicslargescalestructure, coles2025, Uhlemann:2014npa, Uhlemann:2018gzz, Gough_2022, Uhlemann_Kopp_2014},
it is interesting to proceed in the opposite direction, i.e., by using the inverse Madelung transformation, which allows the hydrodynamic equations to be recast into a Schrödinger-like equation. This provides a useful framework for describing cosmic structure formation within the Schrödinger formalism.

In this section, we first revisit the Newtonian approach, where the continuity and Bernoulli equations can be recast into a Schr\"odinger-like equation by combining the density and velocity fields into a single complex scalar field. We then extend this framework to the post-Friedmann regime, deriving the corresponding Schrödinger-like equation in the post-Friedmann approach.

\subsection{The Schr\"odinger-like equation in the Newtonian regime}

In this regime, one can rewrite the continuity and Bernoulli equations in the form of a Schrödinger-like equation by merging the density and velocity fields into a single complex scalar field~\cite{Widrow:1993qq}. This construction relies on expressing the wavefunction in polar form,
\begin{equation}
\psi_N=R_N \exp\left({i\frac{S_N}{\nu}}\right)\ ,
\end{equation}
and performing the Madelung transformation, the Schr\"odinger equation
\begin{equation}\label{Schroedinger_N}
i \nu \frac{\partial \psi_N}{\partial D}=\left[\frac{1}{2}(-i \nu \nabla)^2+V_N\right] \psi_N\ ,
\end{equation}
can be decomposed into an evolution equation for the phase $S$ of the wavefunction and a continuity equation for the squared amplitude $R^2$, in particular, from the real part, one finds
\begin{equation}
\frac{\partial S_N}{\partial D} +\frac{1}{2}( \nabla S_N)^{2} =-V_N+\frac{\nu ^{2}}{2R_N} \nabla ^{2} R_N\ ,
\end{equation}
and the imaginary part gives
\begin{equation}
\frac{\partial R_N^{2}}{\partial D}=-\nabla \left(R_N^{2} \nabla S_N\right) \ .
\end{equation}
These take the same form as the Bernoulli equation, with the effective potential replaced by the potential in the Schr\"odinger equation plus an additional quantum pressure term $-\nu ^{2} \nabla ^{2} R_N/2R_N $, and the continuity equation of Newtonian dynamics, respectively. Therefore, by an inverse procedure, one can combine the Bernoulli and continuity equations in the Newtonian regime into a Schr\"odinger-like equation, with the phase of the wavefunction identified with the velocity potential,
\begin{equation}
\partial ^iS_N= v_N^i\ ,
\end{equation}
the amplitude identified with the square root of the matter density,
\begin{equation}
R^2_N=1+\delta_N\ ,
\end{equation}
and with the potential replaced by the effective potential appearing in the Bernoulli equation,
\begin{equation}
V_N=\mathcal{V}_N\ .
\end{equation}
The subscript $N$ refers to the Newtonian case. The parameter $\nu$ is a real constant that controls the coarse-graining scale of the description, effectively setting the phase-space resolution. In this sense, $\nu$ can be interpreted as encoding the minimal resolved scale in the combined density–velocity description, analogous to a smoothing scale in phase space. 
Precisely, $\nu$ controls the coarse-graining scale, heuristically related to the product of spatial and velocity resolution scales, $\nu \sim \sigma_x \times \sigma_v$.
In analogy with the Schrödinger method for collisionless fluids, $\nu$ plays a role similar to an effective Planck constant, regulating the emergence of small-scale structure and preventing singular behavior such as shell-crossing.
Its value is therefore chosen to match the level of smoothing appropriate for the physical problem under consideration.
Finally, in this Newtonian context, it has been shown that, in the semi-classical limit, i.e., $\nu\rightarrow0$, this wave-mechanical representation of self-gravitating CDM will approach the standard hydrodynamical description \cite{Short_2006a}, while for finite $\nu$, it controls the suppression effect on the gravitational collapse of density perturbations \cite{Coles:2002sj, Short_2006b}.

\subsection{The Schr\"odinger-like equation up to 1PF order}
\label{subsec:Schrodinger_1PF}

One of the main goals of this paper is to extend this framework to the 1PF approximation, namely by combining Eq.~(\ref{Bernoulli}) and Eq.~(\ref{continuity2}) into a single Schrödinger-like equation. As shown in Section II, at 1PF order the velocity field, even in the absence of vorticity, acquires a non-vanishing post-Friedmann vector component, as given in Eq.~(\ref{decomposed V_i}).
This feature obstructs the standard Schr\"odinger-like equation construction, since the presence of a vector component prevents the velocity field from being fully represented as the gradient of a scalar phase. In other words, the Madelung transformation recasts the Schr\"odinger equation into the Newtonian continuity and Bernoulli equations for a curl-free flow; therefore, the Schr\"odinger equation with a standard Hamiltonian operator as in the Newtonian case, i.e., Eq.\ (\ref{Schroedinger_N}), is incapable of reproducing the complete continuity and Euler system for flows with non-vanishing curl.

However, by formal analogy with the dynamics of charged particles in electromagnetic fields, where the Hamiltonian includes a minimal coupling to the magnetic vector potential, see, e.g., \cite{Griffiths_Schroeter_2018}, we can consider the Schr\"odinger equation in a more sophisticated form
\begin{equation}
i \nu \frac{\partial \psi}{\partial D}=\hat{H} \psi\ ,
\end{equation}
where the Hamiltonian operator $\hat{H}$ can be generalized to incorporate a vector potential term, i.e., 
\begin{equation}\label{Hamiltonian}
\hat{H}=\frac{1}{2}(-i \nu \nabla+\mathbf{A})^2+V\ .
\end{equation}
Here, $\mathbf{A}$ is a divergenceless (transverse) vector field satisfying $\nabla\cdot\mathbf{A}=0$.
As we shall show below, with the wavefunction written in the form
\begin{equation}
\psi=R \exp\left({i\frac{S}{\nu}}\right)\ ,
\end{equation}
this Schr\"odinger equation with a generalized Hamiltonian can be recast, via the Madelung transformation, into a set of hydrodynamic equations that is compatible with the 1PF continuity and Bernoulli equations, i.e., Eq.~(\ref{continuity2}) and Eq.~(\ref{Bernoulli}). In this correspondence, $S$ is defined as the scalar velocity potential $\Theta$, while $A^i$ corresponds to the vector component of the velocity field $\varTheta^i$.

\subsubsection{Madelung transformation for a flow with curl}

Following the path of the Madelung transformation, the imaginary part of the Schr\"odinger equation with a generalized Hamiltonian Eq.~(\ref{Hamiltonian}) can be identified as a continuity equation,
\begin{equation}\label{continuity_R}
\frac{\partial R^{2}}{\partial D} +\left[ R^{2}\left( S^{,i} +A^{i}\right)\right]_{,i} =0\ ,
\end{equation}
with a probability flux given by
\begin{equation}
j^i = R^{2}\left( S^{,i} +A^{i}\right) \ .
\end{equation}
The real part of the Schr\"odinger equation yields a Bernoulli-like equation
\begin{equation}\label{Bernoulli_S}
\frac{\partial S}{\partial D} +\frac{1}{2} S_{,k} S^{,k} -\frac{1}{2} \nu ^{2}\frac{\nabla ^{2} R}{R} +A^{k} S_{,k} +\frac{1}{2} A^{2} +V=0\ ,
\end{equation}
with an effective potential given by
\begin{equation}
\mathcal{V}=A^{k} S_{,k} +\frac{1}{2} A^{2} +V\ ,
\end{equation}
except for an extra term  
\begin{equation}
\mathcal{P}=\frac{\nu ^{2} }{2}\frac{\nabla ^{2} R}{R}\,,
\end{equation} 
which is the so-called ``quantum pressure", represents a non-classical contribution driven by spatial gradients of the density, encoding the wave nature of the system. Its role and implementation will be discussed in detail later. Therefore, we have derived the Madelung transformation for the Schrödinger equation with the generalized Hamiltonian given in Eq.~(\ref{Hamiltonian}) and shown that it recasts the equation into the continuity and Euler equations describing a flow with non-vanishing curl. In other words, the set of continuity and Euler equations, where the velocity field is not possible to be represented by a single scalar velocity field, can be transformed into the form of the Schr\"odinger equation for charged particles in an external magnetic field. However, it should be emphasized that the analogy with the case of charged particles in an external magnetic field, formulated via the minimal coupling prescription, is purely formal and does not imply an underlying gauge interaction.

\subsubsection{Inverse Madelung transformation of the 1PF continuity and Bernoulli equations}

To perform the inverse Madelung transformation, we want to define the amplitude $R$ and phase $S$ of the wavefunction, under 1PF approximation of fluid equations.
Since Eq.\ (\ref{Bernoulli}) takes the form of a Bernoulli equation and, except for the quantum pressure term, formally coincides with the evolution equation for the phase obtained from the Madelung transformation of the Schr\"odinger equation, i.e., Eq.\ (\ref{Bernoulli_S}), it is natural to identify the comoving velocity potential $\Theta$ with the phase of the wavefunction, i.e., 
\begin{equation}
S:=\Theta\ .
\end{equation}
This is consistent with the definition of the wavefunction used in the Newtonian regime. However, the usual choice of the amplitude as the square root of the matter density used in the Newtonian regime is not optimal at 1PF order, since the continuity equation for the matter density, Eq.\ (\ref{continuity2}), does not take the Newtonian form, while the continuity equation for the squared amplitude obtained from the Madelung transformation, i.e., Eq.\ (\ref{continuity_R}) retains the form of the continuity equation of a Newtonian fluid. Instead, when expressed in terms of a change of variable in the matter density,
\begin{equation}
\hat{\rho}:=\sqrt{-g}\rho u^0\ ,
\end{equation}
where $g=\det\left(g_{\mu\nu}\right)$ is the determinant of the metric tensor matrix. In terms of this variable, there is
\begin{equation}
\partial_0\hat{\rho}+\frac{1}{ca}\partial_i\left(\hat{\rho}v^i\right)=\partial_0\hat{\rho}+\partial_i\left(\hat{\rho}\frac{u^i}{u^0}\right)=\partial_0\left(\sqrt{-g}\rho u^0\right)+\partial_i\left(\sqrt{-g}\rho u^i\right)=0\ ,
\end{equation}
which follows directly from the covariant conservation law for a pressureless fluid
\begin{equation}
\nabla_\mu\left(\rho u^\mu\right)=\frac{1}{\sqrt{-g}}\partial_\mu\left(\sqrt{-g}\rho u^\mu\right)=0\ .
\end{equation}
Therefore, up to 1PF order, we define
\begin{equation}\label{delta_hat}
\hat{\delta} +1 =( \delta +1)\left[ 1+\frac{1}{c^{2}}\left(\frac{1}{2} a^2H^2f^2D^2 \Theta _{,i} \Theta ^{,i} -3\phi _{G}\right)\right]\ .
\end{equation}
In terms of this variable, the continuity equation beacomes
\begin{equation}\label{continuity newton}
\frac{\partial \hat{\delta}}{\partial D} +V^i \hat{\delta}_{,i} +V^i_{,i}(\hat{\delta} +1) =0\ ,
\end{equation}
which takes the same form as in the Newtonian case \cite{Missinglink,1965ApJ...142.1488C}, and coincides in form with Eq.\ (\ref{continuity_R}).
Note that the definition of $\hat{\delta}$ here is the same as in \cite{Missinglink} (see Eq.~(8.16) therein). 
Therefore, in order to consistently apply the inverse Madelung transformation to the 1PF continuity and Euler equations, we define the amplitude of the wavefunction via 
\begin{gather}
\psi ^{*} \psi =\hat{\delta} +1\ .
\end{gather}
The complete definition of the wavefunction we consider is then
\begin{equation}
\psi=R\exp \left({i\frac{S}{\nu}}\right)=\sqrt{\hat{\delta} +1}\exp \left({i\frac{\Theta}{\nu}}\right)\,.
\end{equation}
By comparing Eq.~(\ref{continuity_R}) with Eq.~(\ref{continuity newton}), one can identify the corresponding velocity fields appearing in the two formulations, i.e., $S^{,i}+A^i=V^i=\Theta^{,i}+\frac{1}{c^2}\varTheta^{i}$. In particular, recalling that the phase $S$ has been defined in terms of the scalar velocity potential $\Theta$, we obtain
\begin{equation}
A^i = \frac{1}{c^2}\varTheta^{i}\ .
\end{equation}

To perform the inverse Madelung transformation, we compute the derivatives of $\psi$. The derivative with respect to the evolution variable $D$ reads
\begin{equation}\label{d psi d D}
\frac{\partial\psi}{\partial D}=\left(\frac{1}{R}\frac{\partial R}{\partial D}+\frac{i}{\nu}\frac{\partial\Theta}{\partial D}\right)\psi\ ,
\end{equation}
and the spatial derivative reads
\begin{gather}
\partial_i\psi=\left(\frac{R_{,i}}{R}+\frac{i}{\nu}\Theta_{,i}\right)\psi\ ,
\end{gather}
while the spatial Laplacian is given by
\begin{equation}
\nabla ^2\psi=\left[\frac{\nabla^2R}{R}-\frac{1}{\nu^2}\Theta_{,k}\Theta^{,k}+\frac{i}{\nu}\left(\frac{2}{R}R_{,k}\Theta^{,k}+\nabla^2\Theta\right)\right]\psi\ .
\end{equation}
Therefore, up to 1PF order, following the generalized Hamiltonian operator Eq.\ (\ref{Hamiltonian}), there is
\begin{equation}\label{Hamiltonian_expanded}
\begin{aligned}
\left(-i\nu\partial_i+\frac{1}{c^2}\varTheta_i\right)^2\psi&=-\left(\nu^2\nabla ^{2} +\frac{1}{c^{2}}2i\nu \varTheta^{i} \partial _{i} \right)\psi\\&=-\left[\nu^2\left(\frac{\nabla^2R}{R}-\frac{1}{\nu^2}\Theta_{,k}\Theta^{,k}+\frac{i}{\nu}\left(\frac{2}{R}R_{,k}\Theta^{,k}+\nabla^2\Theta\right)\right) +\frac{1}{c^{2}}2i\nu \varTheta^{i} \left(\frac{R_{,i}}{R}+\frac{i}{\nu}\Theta_{,i}\right)\right]\psi
\end{aligned}
\end{equation}
Using Eq.\ (\ref{Bernoulli}) and Eq.\ (\ref{continuity newton}) in Eq.\ (\ref{d psi d D}) and Eq.\ (\ref{Hamiltonian_expanded}), we are able to show that the complex wave function $\psi$ satisfies a Schr\"odinger-like equation, i.e.,
\begin{equation}\label{Schrodinger0}
\begin{aligned}
 & i\nu \frac{\partial \psi }{\partial D} -\frac{1}{2}\left( -i\nu \partial _{i} +\frac{1}{c^{2}} \varTheta _{i}\right)^{2} \psi =\left\{\frac{4\pi G\overline{\rho }}{H^{2} f^{2} D} \Theta +\frac{1}{a^{2} H^{2} f^{2} D^{2}}\left[ \phi _{G} +\frac{1}{c^{2}}\left( D_{P} +4\nabla ^{-2}( \phi _{G} \phi _{G,i})^{,i}\right)\right]\right. \\
 & \left.-\frac{1}{c^{2}} \nabla ^{-2} \partial ^{i}\left[ 2\Theta _{,j} \Theta ^{,j} \phi _{G,i} +a^{2} H^{2} f^{2} D^{2} \Theta ^{,j} \Theta ^{,k}( \Theta _{,j} \Theta _{,ki} -\Theta _{,i} \Theta _{,kj}) -\left( 3\frac{\partial \phi _{G}}{\partial D} +\Theta ^{,j} \phi _{G,j} +a^{2} H^{2} fD\Theta _{,j} \Theta ^{,j}\right) \Theta _{,i}\right]\right\} \psi \ .
\end{aligned}
\end{equation}

\subsection{1PF Schr\"odinger equation with an imaginary potential component}

Using the relation given in first line of Eq.\ \ref{Hamiltonian_expanded}, Eq.\ \ref{Schrodinger0} can be written as
\begin{equation}\label{Schrodinger1}
i\nu \frac{\partial \psi }{\partial D} =-\frac{\nu ^{2}}{2} \nabla ^{2} \psi +\frac{\nu ^{2} \nabla ^{2} R}{2R} \psi +\left[\frac{4\pi G\bar{\rho}}{H^2f^2D}\Theta +\frac{1}{a^2H^2f^2D^2} \phi _{G} +\frac{1}{c^{2}}\left( -\frac{i\nu }{R} R_{,i} \varTheta^{i} +\mathcal{A}\right)\right] \psi \ ,
\end{equation}
which is close in form to the standard Schr\"odinger equation but with an imaginary potential term. On the RHS of Eq.\ (\ref{Schrodinger1}), there is an extra term  
$\mathcal{P}=\nu ^{2}\nabla ^{2} R/2R$, commonly referred to as the ``quantum pressure" term. As we have proved before through the Madelung transformation, if we include the quantum pressure as an additional term in the Bernoulli-like equation, it will then not appear in the Schr\"odinger-like equation, i.e., the equation will simplify to
\begin{equation}\label{Schrodinger2}
i\nu \frac{\partial \psi }{\partial D} +\frac{\nu ^{2}}{2} \nabla ^{2} \psi=\left[\frac{4\pi G\bar{\rho}}{H^2f^2D}\Theta +\frac{1}{a^2H^2f^2D^2} \phi _{G} +\frac{1}{c^{2}}\left( -\frac{i\nu }{R} R_{,i} \varTheta^{i} +\mathcal{A}\right)\right] \psi \ .
\end{equation}
Nevertheless, note that, although we can remove the quantum pressure term from the Schr\"odinger-like equation by including it in the Bernoulli-like equation, it plays a rather important role in the Schr\"odinger approach to the structure formation, as it has a function of a regularizing term in the fluid equations. 
This effect has been evaluated and discussed in the Newtonian context \cite{Coles:2002sj, Short_2006a, Short_2006b}. As demonstrated by \cite{Coles:2002sj}, the quantum pressure term functions analogously to a viscous contribution, effectively inhibiting multi-streaming and preventing the formation of density singularities at shell crossing. It is shown in \cite{Short_2006a} that the parameter $\nu$ controls the degree of suppression of gravitational collapse induced by the quantum pressure term. For finite values of $\nu$, the quantum pressure acts prior to shell crossing, moderating the gravitational collapse of the density perturbation. In the semi-classical limit $\nu \to 0$, the quantum pressure term contributes only in regions where particle trajectories intersect, while classical behavior is recovered on large scales \cite{Short_2006b}.

\subsection{Governing equations of the 1PF wave-mechanical formalism}

In order to close the system, we also consider the Einstein field equations. The scalar part of the trace-free part of the spatial components of the Einstein field equations gives
\begin{equation}\label{Dp}
\begin{aligned}
 & \frac{1}{c^{4}}\left\{\frac{2}{3} \nabla ^{2} \nabla ^{2} D_{P} +\left( \phi _{G,i} \phi _{G}^{,j}\right)_{,j}^{,i} -\frac{1}{3} \nabla ^{2}\left( \phi _{G,k} \phi _{G}^{,k}\right)\right\} =-\frac{1}{c^{4}} 4\pi Ga^{4}H^2f^2D^2\overline{\rho }\left[(\delta+1)\left( \Theta _{,i} \Theta ^{,j} -\frac{1}{3} \Theta _{,k} \Theta ^{,k} \delta _{i}^{j}\right)\right]_{,j}^{,i}
\end{aligned}
\end{equation}
Using the temporal component and the trace part of the spatial components of the Einstein field equations, we obtain an equation involving only the potential $\displaystyle \phi _{G}$, sourced by the matter and velocity perturbation
\begin{equation}\label{phi_G}
\begin{aligned}
 & \frac{1}{c^{2}}\frac{2}{3} \nabla ^{2} \phi _{G} +\frac{1}{c^{4}}\left( \frac{\partial ^{2} \phi _{G}}{\partial D^{2}}a^2H^2f^2D^2 +4\pi G\bar\rho a^2D\frac{\partial \phi _{G}}{\partial D} +a^2(2\dot{H}+H^2)\phi _{G} +\frac{2}{3} \nabla ^{2} \phi _{G}^{2} -\frac{3}{2} \phi _{G,i} \phi _{G}^{,i}\right)\\
 & =\frac{1}{c^{2}}\frac{8\pi G}{3} a^{2}\overline{\rho } \delta +\frac{1}{c^{4}} 4\pi Ga^{4}H^2f^2D^2\overline{\rho } (\delta+1)\Theta _{,k} \Theta ^{,k}
\end{aligned}
\end{equation}
We can also obtain a constraint equation for the vector part by applying the operator $\displaystyle \nabla ^{2}$ to the mixed time-space components of the Einstein field equations and $\displaystyle \partial _{i}$ to the scalar part of the equation
\begin{equation}\label{P_i}
\begin{aligned}
 & \frac{1}{c^{3}} \nabla ^{2} \nabla ^{2} P_{i} -\frac{1}{c^{5}}\left\{\nabla ^{2}\left[ 4\left(aHfD\frac{\partial \phi _{G}}{\partial D} \phi _{G,i} +2aHfD \phi _{G}\frac{\partial \phi _{G,i}}{\partial D}\right) +8\dot{a} \phi _{G} \phi _{G,i} +2P_{,i}^{k} \phi _{G,k} -P_{i} \nabla ^{2} \phi _{G} -2P^{k} \phi _{G,ki}\right]\right. \\
 & \ \ \ \ \ \ \ \ \ \ \ \ \ \ \ \ \ \ \ \ \ \ \ \left. -\left[ 4\left(aHfD\frac{\partial \phi _{G}}{\partial D} \phi _{G,j} +2aHfD \phi _{G}\frac{\partial \phi _{G,j}}{\partial D}\right) +8aH\phi _{G} \phi _{G,j} +2P_{,j}^{k} \phi _{G,k} -P_{j} \nabla ^{2} \phi _{G} -2P^{k} \phi _{G,kj}\right]_{,i}^{,j}\right\}\\
 & =-\frac{1}{c^{3}} 16\pi G\overline{\rho } a^{3}HfD\left[ \nabla ^{2}\left( (\delta+1) \Theta _{,i}\right) -\left( (\delta+1) \Theta_{,j}\right)_{,i}^{,j}\right] \\
 &\ \ \ -\frac{1}{c^{5}} 16\pi G\overline{\rho } a^{3}HfD\left\{\nabla ^{2}\left[ (\delta+1)\left[ \varTheta_{i} -\frac{1}{aHfD} P_{i} +\Theta _{,i}\left( a^2H^2f^2D^2 \Theta_{,k} \Theta ^{,k} -4\phi _{G}\right)\right]\right]\right. \\
 &\ \ \ \ \ \ \ \ \ \ \ \ \ \ \ \ \ \ \ \ \ \ \ \ \ \ \ \ \ \ \left.-\left[ (\delta+1)\Theta _{,j}\left(a^2H^2f^2D^2 \Theta _{,k} \Theta ^{,k} -4\phi _{G}\right)\right]_{,i}^{,j}\right\}
\end{aligned}
\end{equation}
The Schr\"odinger-like equation Eq.\ (\ref{Schrodinger2}) together with different components of the Einstein field equations Eq.\ (\ref{Dp}), Eq.\ (\ref{phi_G}), and Eq.\ (\ref{P_i}) form the system of equations that constitutes our 1PF wave-mechanical formalism for a self-gravitating pressureless fluid. 

In our 1PF wave-mechanical formalism for cosmological gravitational instability, the post-Friedmann corrections from the magnetic part of the Weyl tensor, which are responsible for the frame-dragging effect, are encoded in the imaginary potential of the Schr\"odinger-like equation, while in the Newtonian Schr\"odinger approach to the cosmological gravitational instability, the frame-dragging effect is not included. In fact, if we retain terms up to the 0PF order, the resulting system of equations reduces to the Schrödinger–Poisson equations in the Newtonian regime.

\subsection{A curl-free frame for particle velocity fields}

Since the imaginary contribution to the potential originates from the curl of the peculiar velocity field of the particle, it is convenient to perform a coordinate transformation to a frame in which the velocity field is curl-free. In such a frame, the potential does not acquire any imaginary term. 
We therefore introduce the rest frame of a local observer who measures a vanishing curl of the flow. 
The information about the frame-dragging effect is then encoded in the coordinate transformation relating this curl-free frame to the background comoving frame. This construction is analogous in spirit to the orthonormal tetrad frame carried by zero-angular-momentum observers (ZAMO), i.e., the locally nonrotating frames (LNRF) in stationary, axisymmetric, asymptotically flat spacetimes \cite{Bardeen1970_ZAMO, Bardeen1972_LNRF}.

We begin by considering a coordinate transformation $x^\mu\rightarrow\tilde{x}^\mu$, generated by a vector field $\xi^\mu$. The relation between the two coordinates at a given point $x$ is
\begin{equation}
\tilde{x}^{\mu } =x^{\mu } -\xi ^{\mu }( x)\ .
\end{equation}
We specify the transformation vector field as
\begin{equation}
\xi ^{\mu } =\left( 0,\frac{1}{c^{2}} s^{i}\right) \ ;\ s^{i} =s^{,i} +S^{i}\ ,
\end{equation}
where $S^i$ denotes the divergenceless (transverse) vector component, and we also define $s_{i} =\delta _{ij} s^{j} =s_{,i} +S_{i}$.
We have specified that the transformation vector vanishes at 0PF order. This is due to the fact that at the 0PF order, the curl of the velocity field is set to zero as a consequence of the zero vorticity condition. Therefore, the curl-free frame we aim to construct can only differ from the original frame by a transformation that appears at 1PF order. 

As a consequence of coordinate transformation, partial derivatives transform:
\begin{gather}
\frac{\partial }{\partial x^{\mu }} =\frac{\partial \tilde{x}^{\nu }}{\partial x^{\mu }}\frac{\partial }{\partial \tilde{x}^{\nu }} =\left( \delta _{\mu }^{\nu } -\frac{1}{c^{2}} \delta _{i}^{\nu } s_{,\mu }^{i}\right)\frac{\partial }{\partial \tilde{x}^{\nu }}\ ,
\end{gather}
where the Greek indices run over 0,1,2,3 and denote spacetime coordinates, while Latin indices run over 1,2,3 and denote purely spatial coordinates, terms beyond order $1/c^{2}$ have been neglected. Separating the temporal and spatial components yields
\begin{gather}
\frac{\partial }{\partial x^{0}} =\frac{\partial }{\partial \tilde{x}^{0}} -\frac{1}{c^{2}} s_{,0}^{i}\tilde{\partial }_{i} \ \ \ ,\ \ \ \frac{\partial }{\partial x^{i}} =\frac{\partial }{\partial \tilde{x}^{i}} -\frac{1}{c^{2}} s_{,i}^{j}\frac{\partial }{\partial \tilde{x}^{j}}\ .
\end{gather}
The relation between the two metrics evaluated at the same space-time position (not at the same coordinate value), denoted by $x$ and $\tilde{x}$ respectively, is given by
\begin{equation}
\tilde{g}_{\mu \nu }\left(\tilde{x}\right) =\left(\frac{\partial x^{\rho }}{\partial \tilde{x}^{\mu }}\right)\left(\frac{\partial x^{\sigma }}{\partial \tilde{x}^{\nu }}\right) g_{\rho \sigma }( x)\ .
\end{equation}
In the transformed coordinate system, the metric can be parametrized as a sum of terms of different orders of $1/c$, keeping terms up to $O(1/c^4)$, we can write different components of the metric tensor as
\begin{gather}
\tilde{g}_{00} =-1-\frac{2}{c^{2}}\tilde{\phi }_{G} -\frac{2}{c^{4}}\left(\tilde{\phi }_{G}^{2} +\tilde{D}_{P}\right)\ ,\\
\tilde{g}_{0i} =-\frac{1}{c^{3}} a\left(\tilde{\partial }_{i}\tilde{p} +\tilde{P}_{i}\right) \ ,\\
\tilde{g}_{ij} =a^{2}\left\{\left[ 1-\frac{2}{c^{2}}\tilde{\phi }_{G} +\frac{2}{c^{4}}\left(\tilde{\phi }_{G}^{2} +\tilde{D}_{P}\right)\right] \delta _{ij} +\frac{1}{c^{2}} 2\tilde{\partial }_{i}\tilde{\partial }_{j}\tilde{\gamma } +\frac{1}{c^{2}} 2\tilde{\partial }_{( j}\tilde{\omega }_{i)} +\frac{1}{c^{4}}\tilde{h}_{ij}\right\}\ ,
\end{gather}
where $\tilde{P}_i$ and $\tilde{\omega}_i$ are both divergenceless (transverse) 
satisfying $\tilde{\partial}^i\tilde{P}_i=0$ and $\tilde{\partial}^i\tilde{\omega}_i=0$.
Hence, the relation between the metric components in the $\tilde{x}^\mu$ frame and the $x^\mu$ frame is given by
\begin{gather}
\frac{1}{c^{2}}\tilde{\phi }_{G}=\frac{1}{c^{2}} \phi _{G} \ \ \ \ ,\ \ \ \ \frac{1}{c^{4}}\tilde{D}_{P} =\frac{1}{c^{4}}D_{P}\ , \notag\\
\frac{1}{c^{3}}\tilde{p} =-\frac{1}{c^{3}} a\frac{\partial s}{\partial t}\ \ \ \ ,\ \ \ \ \frac{1}{c^{3}}\tilde{P}_{i} =\frac{1}{c^{3}}\left( P_{i} -a\frac{\partial S_{i}}{\partial t}\right)\ ,\\
\frac{1}{c^{2}}\tilde{\gamma } =\frac{1}{c^{2}}s\ \ \ \ ,\ \ \ \ \frac{1}{c^{2}}\tilde{\omega }_{i} =\frac{1}{c^{2}} S_{i} \ .\notag
\end{gather}
Note that the left-hand side of each equation is evaluated at $\tilde{x}^\mu$, while the right-hand side at $x^\mu\left(\tilde{x}\right)$. The choice of time slicing in the new coordinate system is equivalent to that of the Poisson gauge; therefore, only the variables that depend on the spatial choice are affected.

To derive the hydrodynamical equations in the tilded frame, we need to express the metric and matter variables appearing in Eq.\ (\ref{continuity1}) and Eq.\ (\ref{Euler1}) as functions of the corresponding variables in the tilded frame. We have already derived the relation between the metric variables in the two frames; now, we turn our attention to the relation between matter variables. The energy density remains unchanged,
\begin{equation}
\tilde{\rho }\left(\tilde{x}\right) =-\frac{1}{c^{2}} T_{\mu }^{\mu } =\rho ( x) \ ,
\end{equation}
and the density contrast is also unchanged since the background energy density depends only on time and we are not performing a transformation of the time coordinate
\begin{equation}
\delta \left( x\right) =\frac{\rho \left( x^{\mu }\right) -\overline{\rho }\left( x^{0}\right)}{\overline{\rho }\left( x^{0}\right)} =\frac{\tilde{\rho }\left(\tilde{x}^{\mu }\right) -\overline{\rho }\left(\tilde{x}^{0}\right)}{\overline{\rho }\left(\tilde{x}^{0}\right)} =\tilde{\delta }\left(\tilde{x}\right)\ .
\end{equation}
Consequently,
\begin{equation}
\tilde{\hat{\delta}}+1 = (\tilde \delta +1)\left[ 1+\frac{1}{c^{2}}\left(\frac{1}{2} \dot{a}^{2}f^2D^2\tilde \Theta _{,i}\tilde \Theta ^{,i} -3\tilde \phi _{G}\right)\right]=\hat{\delta}+1\ .
\end{equation}
While the physical peculiar velocity is changed since the four-velocity would change under the coordinate transformation, following
\begin{equation}
\displaystyle u^{\mu }( x) =\left(\frac{\partial x^{\mu }}{\partial \tilde{x}^{\rho }}\right)\tilde{u}^{\rho }\left(\tilde{x}\right)\ \rightarrow \ u^{0}( x) =\tilde{u}^{0}\left(\tilde{x}\right) \ ;\ u^{i}( x) =\tilde{u}^{i}\left(\tilde{x}\right) +\frac{1}{c^{2}} s_{,\rho }^{i}\tilde{u}^{\rho }\left(\tilde{x}\right)\ .
\end{equation}
Thus, remembering the relation between the physical peculiar velocity $v^i$ and the four-velocity $u^i$, $v^{i}( x) =acu^{i}/u^{0}$, we find relation between the velocity fields in the two different frames
\begin{gather}
v^{i}( x) =ac\frac{\tilde{u}^{i}}{\tilde{u}^{0}} +\frac{1}{c^{2}} ac\frac{s_{,\rho }^{i}\tilde{u}^{\rho }}{\tilde{u}^{0}} =\tilde{v}^{i}\left(\tilde{x}\right) +\frac{1}{c^{2}}\left( a\dot{s}^{i} +s_{,j}^{i} v^{j}\right)\ .
\end{gather}
Therefore, we can write down the continuity and Euler equations in the new coordinates up to 1PF order, which read
\begin{gather}
\dot{\tilde{\delta }} +\frac{\tilde{v}^{i}\tilde{\delta }_{,i}}{a} +\frac{\left( 1+\tilde{\delta }\right)\tilde{v}^{i}{}_{,i}}{a} +\frac{1}{c^{2}}\left( 1+\tilde{\delta }\right)\left[ -\frac{\dot a}{a}\tilde{v}^{2}  -3\dot{\tilde{\phi }}_{G} -4\frac{\tilde{v}^{i}\tilde{\phi }_{G,i}}{a}+\nabla ^{2}\dot{s} +\frac{\nabla ^{2} s_{,i}\tilde{v}^{i}}{a}\right] =0\ ,
\end{gather}
and
\begin{equation}
\begin{aligned}
 & \dot{\tilde{v}}_{i} +\frac{\dot{a}}{a}\tilde{v}_{i} +\frac{1}{a}\tilde{v}^{j}\tilde{v}_{i,j} +\frac{1}{a}\tilde{\phi }_{G,i}\\
 & +\frac{1}{c^{2}}\left[\frac{1}{a}\tilde{\phi }_{G,i}\left( 4\tilde{\phi }_{G} +\tilde{v}_{k}\tilde{v}^{k}\right) -3\tilde{v}_{i}\left(\dot{\tilde{\phi }}_{G} +\frac{\tilde{v}^{j}\tilde{\phi }_{G,j}}{a}\right) +\frac{1}{a}\tilde{D}_{P,i} -\frac{1}{a}\tilde{v}_{i}\tilde{v}_{j}\tilde{\phi }_{G}^{,j} -\frac{\dot{a}}{a}\tilde{v}_{k}\tilde{v}^{k}\tilde{v}_{i}\right. \\
 & \left. +\frac{1}{a}\left(\tilde{P}_{j} +a\dot{S}_{j}\right)_{,i}\tilde{v}^{j} -\left(\tilde{P}_{i} -a\dot{s}_{,i} -s_{i,j}\tilde{v}^{j}\right)^{\dot{}} -\frac{1}{a}\tilde{v}^{j}\left(\tilde{P}_{i} -a\dot{s}_{,i} +s_{i,k}\tilde{v}^{k}\right)_{,j} -\frac{\dot{a}}{a}\left(\tilde{P}_{i} -a\dot{s}_{,i} +s_{i,j} v^{j}\right) -\frac{1}{a} s_{,i}^{k}\tilde{\phi }_{G,k}\right] =0\ .
\end{aligned}
\end{equation}
Note that, from now on, commas in equations will denote partial derivatives with respect to the spatial coordinates $\tilde{x}^i$.
The continuity equation written in terms of $\tilde{\hat{\delta}}$ is then
\begin{equation}
\dot{\tilde{\hat \delta }} +\frac{\tilde{v}^{i}\tilde{\hat \delta }_{,i}}{a} +\frac{\left( 1+\tilde{\hat \delta }\right)\tilde{v}^{i}{}_{,i}}{a} +\frac{1}{c^{2}}\left( 1+\tilde{\hat \delta }\right)\left(\nabla ^{2}\dot{s} +\frac{\nabla ^{2} s_{,i}\tilde{v}^{i}}{a}\right)=0\ .
\end{equation}

So far, the transformation is rather general; we have not yet determined the transformation vector $\displaystyle s^{i} =s^{,i} +S^{i}$. Since we are aiming at a curl-free frame for the velocity field, we require 
\begin{equation}
{\displaystyle \tilde{v}_{i} \equiv \tilde{\partial }_{i}\tilde{\theta }}\ \rightarrow\ {\displaystyle \tilde{\partial }_{j}\tilde{v}_{i} -\tilde{\partial }_{i}\tilde{v}_{j} =0}\ ,
\end{equation}
which implies a dynamical equation of $s^i$,
\begin{equation}
-\frac{1}{c^{2}} s_{,j}^{k}\tilde{\theta }_{,ik} +\frac{1}{c^{2}} s_{,i}^{k}\tilde{\theta }_{,jk} +\frac{1}{c^{2}}\left( a\dot{s}_{i} +s_{i,k}\tilde{\theta }^{,k}\right)_{,j} -\frac{1}{c^{2}}\left( a\dot{s}_{j} +s_{j,k}\tilde{\theta }^{,k}\right)_{,i} =\frac{1}{c^{2}} ( \vartheta_{i,j} - \vartheta_{j,i} )\ .
\end{equation}
Due to the anti-symmetry of the equation, the vector field $s^i$ has to satisfy
\begin{equation}\label{transformation,Pi,ij}
-\frac{1}{c^{2}} s_{,j}^{k}\tilde{\theta }_{,ik} +\frac{1}{c^{2}}\left( a\dot{s}_{i} +s_{i,k}\tilde{\theta }^{,k}\right)_{,j} =\frac{1}{c^{2}} \vartheta_{i,j} +\Pi _{( ij)}\ ,
\end{equation}
where $\displaystyle \Pi _{( ij)}$ is some arbitrary symmetric tensor. The trace of Eq.\,(\ref{transformation,Pi,ij}) indicates 
\begin{gather}
\frac{1}{c^{2}}\left( a\tilde\nabla ^{2}\dot{s} +\tilde\nabla ^{2} s_{,k}\tilde{\theta }^{,k}\right) =\Pi _{i}^{i}\ \ 
\rightarrow \ \frac{1}{c^{2}}\frac{d}{dt} \tilde\nabla ^{2} s=\Pi _{i}^{i}\ ,
\end{gather}
which will simplify the continuity equation to the same form as the one in the $\displaystyle x^{\mu }$ frame if $\displaystyle \Pi _{( ij)} =0$, and, in this case, after the inverse Madelung transformation, we should not have any imaginary potential in the Schr\"odinger equation. Therefore, we require
\begin{equation}\label{transformation,ij}
-\frac{1}{c^{2}} s_{,j}^{k}\tilde{\theta }_{,ik} +\frac{1}{c^{2}}\left( a\dot{s}_{i} +s_{i,k}\tilde{\theta }^{,k}\right)_{,j} =\frac{1}{c^{2}} \vartheta_{i,j}\ ,
\end{equation}
operating $\tilde \partial ^{j}$ and $\tilde \nabla ^{-2}$on Eq.\,(\ref{transformation,ij}) we find the equation that determines the coordinate transformation
\begin{equation}\label{transformation,i}
\frac{1}{c^{2}}\left[ a\dot{s}_{i} +s_{i,k}\tilde{\theta }^{,k} -\tilde\nabla ^{-2}\left( s_{,j}^{k}\tilde{\theta }_{,ik}\right)^{,j}\right] =\frac{1}{c^{2}} \vartheta_{i}\ .
\end{equation}
Thus, a vector field $s_i(t, x_i)$ that satisfies Eq.\,(\ref{transformation,i}) 
defines a coordinate transformation leading to a curl-free velocity field. Under this transformation, the continuity and Euler equations become
\begin{equation}
\dot{\tilde{\hat \delta }} +\frac{\tilde{\theta}^{,i}\tilde{\hat \delta }_{,i}}{a} +\frac{\left( 1+\tilde{\hat \delta }\right)\tilde{\nabla}^2\tilde{\theta}}{a}=0\ ,
\end{equation}
and
\begin{equation}
\begin{aligned}
 & \dot{\tilde{\theta }}_{,i} +\frac{\dot{a}}{a}\tilde{\theta }_{,i} +\frac{1}{a}\tilde{\theta }^{,j}\tilde{\theta }_{,ij} +\frac{1}{a}\tilde{\phi }_{G,i}\\
 & +\frac{1}{c^{2}}\left[\frac{1}{a}\tilde{\phi }_{G,i}\left( 4\tilde{\phi }_{G} +\tilde{\theta }_{,k}\tilde{\theta }^{,k}\right) -3\tilde{\theta }_{,i}\dot{\tilde{\phi }}_{G} +\frac{1}{a}\tilde{D}_{P,i} -\frac{4}{a}\tilde{\theta }_{,i}\tilde{\theta }_{,j}\tilde{\phi }_{G}^{,j} -\frac{\dot{a}}{a}\tilde{\theta }_{,k}\tilde{\theta }^{,k}\tilde{\theta }_{,i}\right. \\
 & \left. +\frac{1}{a}\left(\tilde{P}_{j} +a\dot{S}_{j}\right)_{,i}\tilde{\theta }^{,j} -\left(\tilde{P}_{i} -a\dot{s}_{,i} -s_{i,j}\tilde{\theta }^{,j}\right)^{\dot{}} -\frac{1}{a}\tilde{\theta }^{,j}\left(\tilde{P}_{i} -a\dot{s}_{,i} -s_{i,k}\tilde{\theta }^{,k}\right)_{,j} -\frac{\dot{a}}{a}\left(\tilde{P}_{i} -a\dot{s}_{,i} -s_{i,j}\tilde{\theta }^{,j}\right) -\frac{1}{a} s_{,i}^{k}\tilde{\phi }_{G,k}\right] =0\ .
\end{aligned}
\end{equation}
By performing the inverse Madelung transformation, we obtain the Schrödinger-like equation in the velocity curl-free frame
\begin{equation}\label{tilde Schroedinger}
i\nu \frac{\partial\tilde{\psi }}{\partial D} =-\frac{\nu ^{2}}{2} \tilde \nabla ^{2} \tilde\psi +\frac{\nu ^{2} \tilde \nabla ^{2}\tilde R}{2\tilde R}\tilde \psi +\left(4\pi G\bar{\rho}\frac{a^2}{\dot{a}^2f^2D}\tilde \Theta+\frac{1}{\dot{a}^{2}f^2D^2} \tilde \phi _{G} +\frac{1}{c^{2}}\tilde{\mathcal{A}}\right)\tilde \psi \ ,
\end{equation}
where
\begin{equation}
\begin{aligned}
\frac{1}{c^{2}}\tilde{\mathcal{A}} = & \frac{1}{c^{2}} \tilde\nabla ^{-2} \tilde\partial ^{i}\left[\tilde{\phi }_{G,i}\left(\frac{4}{\dot{a}^{2}f^2D^2}\tilde{\phi }_{G} +\tilde{\Theta }_{,k}\tilde{\Theta }^{,k}\right) -3\tilde{\Theta }_{,i}\frac{\partial \tilde{\phi }_{G}}{\partial D} +\frac{1}{\dot{a}^{2}f^2D^2}\tilde{D}_{P,i} -4\tilde{\Theta }_{,i}\tilde{\Theta }_{,j}\tilde{\phi }_{G}^{,j} -\dot{a}^2fD\tilde{\Theta }_{,k}\tilde{\Theta }^{,k}\tilde{\Theta }_{,i}\right. \\
 & \left. \ \ \ \ \ \ \ \ \ \ \ \ \ +\left(\frac{1}{\dot{a}fD}\tilde{P}_{j} +\frac{\partial S_{j}}{\partial D}\right)_{,i}\tilde{\Theta }^{,j} +\frac{\partial }{\partial D}\left(\frac{\partial s_{,i}}{\partial D} +s_{i,j}\tilde{\Theta }^{,j}\right) +4\pi G\bar{\rho}\frac{a^2}{\dot{a}^2f^2D}\left(\frac{\partial s_{,i}}{\partial D} +s_{i,j}\tilde{\Theta }^{,j}\right)\right. \\
 & \left. \ \ \ \ \ \ \ \ \ \ \ \ \ -\tilde{\Theta }^{,j}\left(\frac{1}{\dot{a}fD}\tilde{P}_{i} -\frac{\partial s_{,i}}{\partial D} - s_{i,k}\tilde{\Theta }^{,k}\right)_{,j} -\frac{1}{\dot{a}^{2}f^2D^2} s_{,i}^{k}\tilde{\phi }_{G,k}\right]\ .
\end{aligned}
\end{equation}
We have used the linear growth factor $D(t)$ as the evolution variable and have defined
\begin{gather}
\tilde\psi \equiv \tilde Re^{\frac{i\tilde\Theta }{\nu }} \ ;\ \tilde\psi ^{*} \tilde\psi =\tilde R^{2} \equiv \tilde{\hat{\delta}}+1 = (\tilde \delta +1)\left[ 1+\frac{1}{c^{2}}\left(\frac{1}{2} \dot{a}^{2}f^2D^2\tilde \Theta _{,i}\tilde \Theta ^{,i} -3\tilde \phi _{G}\right)\right]\ .
\end{gather}

In this velocity curl-free frame, the Schrödinger-like equation Eq.\ (\ref{tilde Schroedinger}) no longer contains the imaginary potential term, which is associated with the frame-dragging effect. We thus arrive at a frame in which the velocity field is purely described by a scalar potential, leading to a real potential in the Schr\"odinger-like equation, more closely resembling the standard Schr\"odinger equation.
This construction bears a close resemblance to the philosophy of the Lagrangian framework of hydrodynamics, e.g., \cite{Buchert1993, Buchert1994, Bouchet1995, Catelan:1994ze}. In both approaches, one seeks a coordinate system that follows, or partially follows, the fluid motion and absorbs part of the nonlinear dynamics into a displacement field. In this sense, the transformation vector $s^i$ effectively plays the role of a displacement field.
There are two main benefits obtained from this coordinate transformation: one is that the frame-dragging effect can be extracted and studied separately via the transformation vector; another one is that the potential is real and thus more suitable for finding the solution.

\section{Conclusions}

In this work, we have extended the Schr\"odinger method for the evolution of CDM beyond the Newtonian regime by constructing a general relativistic wave-mechanical framework within the post-Friedmann approximation \cite{Missinglink}, which is a nonlinear relativistic framework for structure formation, in which 0PF order corresponds to the Newtonian regime and 1PF order terms introduces general relativistic corrections to the equations. Starting from the Einstein field equations and the conservation of the energy–momentum tensor for a pressureless fluid, we derived the 1PF continuity and Euler equations, Eq.~(\ref{continuity3}) and Eq.~(\ref{Euler3}). Equipped with the continuity and Euler equations up to 1PF order, we encountered a difficulty that arises when entering the nonlinear regime beyond Newtonian dynamics. That is, linear perturbation theory implies a curl-free velocity field, which is guaranteed by the Kelvin circulation theorem, but, in our nonlinear post-Friedmann approximation, despite the conservation of vorticity remaining valid, the vanishing of the vorticity does not imply a curl-free velocity field from 1PF order onwards. Specifically, the velocity field with zero vorticity is curl-free to 0PF order but not to 1PF order, i.e., it can be decomposed into a scalar (curl-free) part that is non-zero to all orders and a vector (divergence-free) part that is zero at 0PF order and non-zero from 1PF order onwards. The decomposition of the velocity field is given by Eq.~(\ref{decomposed V_i}). Applying this decomposition to the velocity field and performing the inverse Madelung transformation \cite{Widrow:1993qq, Madelung:1927ksh}, we obtained the Schr\"odinger-like equation, Eq.~(\ref{Schrodinger2}), which includes general relativistic corrections in the potential. Together with the Einstein field equations up to 1PF order, Eq.~(\ref{Dp}), Eq.~(\ref{phi_G}), and Eq.~(\ref{P_i}), we obtain a nonlinear post-Friedmann wave-mechanical framework for structure formation, which is consistent with the standard Schrödinger–Poisson system (c.f. \cite{Coles:2002sj, Coles:2001fw, Short_2006a, Short_2006b, Gallagher_2022, Szapudi_2003, Johnston2010, coles2002wavemechanicslargescalestructure, coles2025, Uhlemann:2014npa, Uhlemann:2018gzz, Gough_2022, Uhlemann_Kopp_2014}) at leading order.

The key result we found is that the 1PF vector part of the velocity field causes a complex potential in the Schr\"odinger-like equation we obtained by performing the inverse Madelung transformation to the 1PF continuity and Euler equations. It makes the Schr\"odinger-like equation similar to the Schr\"odinger equation of a charged particle moving in an external magnetic field. The vector part of the velocity field acts effectively analogous to the magnetic vector potential, relating to a frame-dragging effect. Guaranteed by the diffeomorphism covariance of GR, we then defined a 1PF order coordinated transformation, in which the Schr\"odinger-like equation does not contain any imaginary part in the potential. The transformation vector satisfies Eq.~(\ref{transformation,i}).
This velocity field curl-free frame simplifies the form of the potential and therefore makes the Schr\"odinger-like equation closer to the form of the standard Schr\"odinger equation, Eq.~(\ref{tilde Schroedinger}), while the frame-dragging contribution is isolated and encoded in the transformation vector field. In a following paper in preparation, we will explore the solution to the transformation vector and find an approximate propagator for the Schr\"odinger-like equation.

In summary, we have constructed a nonlinear 1PF wave-mechanical framework for cosmological structure formation and provided a frame of reference in which we can separate the frame-dragging effect that naturally appears under an approximation beyond the Newtonian regime.

\begin{acknowledgments}
We thank Cora Uhlemann for useful discussions.

YDH acknowledges a fellowship from the China Scholarship Council (CSC). YDH, DB, and SM acknowledge financial support from the INFN InDark initiative. This work is supported in
part by the MUR Departments of Excellence grant “Quantum Frontiers” of the Physics and
Astronomy Department of Padova.
\end{acknowledgments}

\appendix

\section{1PF Weyl tensor}
The Weyl tensor can be decomposed into an electric and a magnetic part, as can be seen in the following. In this respect, the dynamics are analogous to the electromagnetic dynamics. This analogy can be demonstrated by writing the Bianchi identities,
\begin{equation}
R^\mu{ }_{\nu[\sigma \gamma ; \rho]}=0\ ,
\end{equation}
in terms of the Weyl tensor, i.e., 
\begin{equation}
C^{\mu\nu\sigma\rho}{ }_{;\rho}=R^{\sigma[\mu;\nu]}-\frac{1}{6}g^{\sigma[\mu}R^{;\nu]}\ .    
\end{equation}
The Weyl tensor can be decomposed into its electric part $E_{\mu\nu}$ and magnetic part $H_{\mu\nu}$ relative to a timelike congruence $u^\mu$, with both tensors being fully projected onto the spatial hypersurfaces orthogonal to $u^\mu$,
\begin{equation}
E_{\mu \nu}=C_{\mu \sigma \nu \gamma} u^\sigma u^\gamma, \quad H_{\mu \nu}=\frac{1}{2} \eta_{\mu \sigma \gamma} C^{\sigma \gamma}{ }_{\nu \rho} u^\rho\ .
\end{equation}
In terms of $E_{\mu\nu}$ and $H_{\mu\nu}$, the Bianchi identities take a form similar to Maxwell's equations, see, \cite{Ellis:1971pg, Ellisbook} for the full sets of equations. In the context of cosmological gravitational instability, the governing dynamical equations contain both the electric and magnetic parts of the Weyl tensor, see, e.g., \cite{Kofman_1995, Matarrese_1994, Bertschinger:1994nc}. The electric part is associated with a tidal acceleration, whereas the magnetic part is associated with the differential frame-dragging, and the frame-dragging vector potential is closely related to the transverse part of the fluid’s velocity, 
see, e.g., \cite{Estabrook1964, Rampf_2014, Nichols_2011}.
To see a direct relation between $H_{ij}$, and the physical velocity field $v^i$, we start from the Ricci identities for the four-velocity vector $u^{\mu}$, i.e.,
\begin{equation}\label{Ricci identities}
u_{\alpha;\nu;\mu}-u_{\alpha;\mu;\nu}=R_{\alpha\beta\mu\nu}u^{\beta}\ .
\end{equation}
Multiplying Eq.~(\ref{Ricci identities}) by $\eta_{\alpha\mu\nu}$, symmetrizing over the free indices, and considering a pressureless dust component with zero vorticity, we obtain the relation (for a more general relation including vorticity and pressure, see \cite{Ellisbook, Ellis:1971pg})
\begin{equation}
H_{\alpha \beta }  =\eta _{\ \ ( \alpha }^{\rho \nu } h^{\mu}_{\beta )}\sigma _{\mu \nu ;\rho }\ ,
\end{equation}
where $\sigma_{\mu\nu}$ is the shear tensor, given by
\begin{equation}
\sigma_{\mu\nu}=\frac{1}{2}\left( h_{\mu }^{\gamma } h_{\nu }^{\sigma } + h_{\nu }^{\gamma } h_{\mu }^{\sigma }\right) u_{\gamma ;\sigma }-\frac{1}{3} u^{\sigma}_{;\sigma }h_{\mu\nu}\ .
\end{equation}
Therefore, for a pressureless dust component with zero vorticity, the magnetic part of the Weyl tensor is given by
\begin{equation}
H_{\alpha \beta }=\eta _{\ \ (\alpha }^{\rho \nu } h_{\beta )}^{\mu }\left( u_{\mu ;\nu } -\frac{1}{3} h_{\mu \nu } u_{;\sigma }^{\sigma }\right)_{;\rho }\ .
\end{equation}
The spatial components $H_{ij}$ are then given by
\begin{equation}
H_{ij}=\eta _{\ \ (i}^{\rho \nu } h_{0j)} u_{;\nu ;\rho }^{0} -\frac{1}{3} \eta _{\ \ (i}^{\rho \nu } h_{j)}^{\mu }\left( h_{\mu \nu } u_{;0}^{0}\right)_{;\rho } +\frac{1}{ac}\left[ \eta _{\ \ (i}^{\rho \nu } h_{kj)}\left( u^{0} v^{k}\right)_{;\nu ;\rho } -\frac{1}{3} \eta _{\ \ (i}^{\rho \nu } h_{j)}^{\mu }\left( h_{\mu \nu }\left( u^{0} v^{k}\right)_{;k}\right)_{;\rho }\right]\ ,
\end{equation}
where we have applied $u^i=(1/ac)u^0v^i$ and the decomposition of $v^i$ with vanishing vorticity, $v^i=\theta^{,i}+1/c^2\vartheta^i$. The magnetic part of the Weyl tensor has a clear association to the curl of the physical velocity field.
Under the PF approximation, the first nonvanishing components of $H_{\mu\nu}$ and $E_{\mu\nu}$ are given by
\begin{equation}
H_{ij}=\frac{1}{2c^3}\left(P_{k,l(i}\varepsilon_{j)}^{\ \ kl}-4v_k\phi_{G,l(i}\varepsilon_{j)}^{\ \ kl}\right)\ ,
\end{equation}
\begin{equation}
E_{ij} =\frac{1}{c^{2}}\left( \phi _{G,ij} -\frac{1}{3} \delta _{ij} \nabla ^{2} \phi _{G}\right)\ .
\end{equation}
Using the curl of $v^i$ given in Eq.~(\ref{theta}), there is
\begin{equation}
\frac{1}{c^{2}} \vartheta _{k,l( i} \varepsilon _{j)}^{\ \ kl} =\frac{1}{c^{2}}\left\{2c^{3} H_{ij} -\theta _{,k( i}\left( -3\phi _{G} +\frac{1}{2} \theta _{,a} \theta ^{,a}\right)_{,l} \varepsilon _{j)}^{\ \ kl} -\theta _{,k}\left( -7\phi _{G} +\frac{1}{2} \theta _{,a} \theta ^{,a}\right)_{,l( i} \varepsilon _{j)}^{\ \ kl}\right\}\ .
\end{equation}

\bibliography{reference}

@article{Missinglink,
  title = {Missing link: A nonlinear post-Friedmann framework for small and large scales},
  author = {Milillo, Irene and Bertacca, Daniele and Bruni, Marco and Maselli, Andrea},
  journal = {Phys. Rev. D},
  volume = {92},
  issue = {2},
  pages = {023519},
  numpages = {19},
  year = {2015},
  month = {Jul},
  publisher = {American Physical Society},
  doi = {10.1103/PhysRevD.92.023519},
  url = {https://link.aps.org/doi/10.1103/PhysRevD.92.023519}
}

@article{Bernardeau_2002,
   title={Large-scale structure of the Universe and cosmological perturbation theory},
   volume={367},
   ISSN={0370-1573},
   url={http://dx.doi.org/10.1016/S0370-1573(02)00135-7},
   DOI={10.1016/s0370-1573(02)00135-7},
   number={1–3},
   journal={Physics Reports},
   publisher={Elsevier BV},
   author={Bernardeau, F. and Colombi, S. and Gaztañaga, E. and Scoccimarro, R.},
   year={2002},
   month=sep, pages={1–248} }

@article{Sahni_1995,
   title={Approximation methods for non-linear gravitational clustering},
   volume={262},
   ISSN={0370-1573},
   url={http://dx.doi.org/10.1016/0370-1573(95)00014-8},
   DOI={10.1016/0370-1573(95)00014-8},
   number={1–2},
   journal={Physics Reports},
   publisher={Elsevier BV},
   author={Sahni, V},
   year={1995},
   month=nov, pages={1–135} }

@BOOK{Peebles1980,
       author = {{Peebles}, P.~J.~E.},
        title = "{The large-scale structure of the universe}",
         year = 1980,
       adsurl = {https://ui.adsabs.harvard.edu/abs/1980lssu.book.....P}
}

@article{Catelan1994,
    author = "Catelan, Paolo and Lucchin, Francesco and Matarrese, Sabino and Moscardini, Lauro",
    title = "{Eulerian perturbation theory in nonflat universes: Second order approximation}",
    eprint = "astro-ph/9411066",
    archivePrefix = "arXiv",
    reportNumber = "COSMO-PD-22",
    doi = "10.1093/mnras/276.1.39",
    journal = "Mon. Not. Roy. Astron. Soc.",
    volume = "276",
    pages = "39",
    year = "1995"
}

@article{Zeldovich,
    author = "Zeldovich, Ya. B.",
    title = "{Gravitational instability: An Approximate theory for large density perturbations}",
    journal = "Astron. Astrophys.",
    volume = "5",
    pages = "84--89",
    year = "1970"
}

@article{Buchert1993,
    author = "Buchert, T.",
    title = "{Lagrangian perturbation theory: A key model for large scale structure}",
    journal = "Astron. Astrophys.",
    volume = "267",
    pages = "L51--L54",
    year = "1993"
}

@article{Buchert1994,
    author = "Buchert, Thomas",
    title = "{Lagrangian theory of gravitational instability of Friedman-Lemaitre cosmologies: Generic third order model for nonlinear clustering}",
    eprint = "astro-ph/9309055",
    archivePrefix = "arXiv",
    doi = "10.1093/mnras/267.4.811",
    journal = "Mon. Not. Roy. Astron. Soc.",
    volume = "267",
    pages = "811--820",
    year = "1994"
}

@article{Bouchet1995,
    author = "Bouchet, F. R. and Colombi, S. and Hivon, E. and Juszkiewicz, R.",
    title = "{Perturbative Lagrangian approach to gravitational instability}",
    eprint = "astro-ph/9406013",
    archivePrefix = "arXiv",
    reportNumber = "FERMILAB-PUB-94-168-A",
    journal = "Astron. Astrophys.",
    volume = "296",
    pages = "575",
    year = "1995"
}

@article{Catelan:1994ze,
    author = "Catelan, Paolo",
    title = "{Lagrangian dynamics in nonflat universes and nonlinear gravitational evolution}",
    eprint = "astro-ph/9406016",
    archivePrefix = "arXiv",
    reportNumber = "OUAST-94-19",
    doi = "10.1093/mnras/276.1.115",
    journal = "Mon. Not. Roy. Astron. Soc.",
    volume = "276",
    pages = "115",
    year = "1995"
}

@article{colesTZ,
author = {Coles, Peter and Melott, Adrian and Shandarin, Sergei},
year = {1993},
month = {03},
pages = {},
title = {Testing approximations for non-linear gravitational clustering},
volume = {260},
journal = {Monthly Notices of the Royal Astronomical Society},
doi = {10.1093/mnras/260.4.765}
}

@article{Melott:1994ah,
    author = "Melott, A. L. and Buchert, T. and Weiss, A. G.",
    title = {Testing higher order Lagrangian perturbation theory against numerical simulations. 2: Hierarchical models},
    eprint = "astro-ph/9404018",
    archivePrefix = "arXiv",
    journal = "Astron. Astrophys.",
    volume = "294",
    pages = "345--365",
    year = "1995"
}

@article{Buchert:1997dr,
    author = "Buchert, Thomas and Dominguez, Alvaro",
    title = "{Modeling multistream flow in collisionless matter: approximations for large scale structure beyond shell crossing}",
    eprint = "astro-ph/9702139",
    archivePrefix = "arXiv",
    journal = "Astron. Astrophys.",
    volume = "335",
    pages = "395--402",
    year = "1998"
}

@article{Gurbatov:1989az,
    author = "Gurbatov, S. N. and Saichev, A. I. and Shandarin, S. F.",
    title = "{The large-scale structure of the universe in the frame of the model equation of non-linear diffusion}",
    journal = "Mon. Not. Roy. Astron. Soc.",
    volume = "236",
    pages = "385--402",
    year = "1989"
}

@article{Carrasco_2012,
   title={The effective field theory of cosmological large scale structures},
   volume={2012},
   ISSN={1029-8479},
   url={http://dx.doi.org/10.1007/JHEP09(2012)082},
   DOI={10.1007/jhep09(2012)082},
   number={9},
   journal={Journal of High Energy Physics},
   publisher={Springer Science and Business Media LLC},
   author={Carrasco, John Joseph M. and Hertzberg, Mark P. and Senatore, Leonardo},
   year={2012},
   month=sep }

@article{Carrasco:2013mua,
    author = "Carrasco, John Joseph M. and Foreman, Simon and Green, Daniel and Senatore, Leonardo",
    title = "{The Effective Field Theory of Large Scale Structures at Two Loops}",
    eprint = "1310.0464",
    archivePrefix = "arXiv",
    primaryClass = "astro-ph.CO",
    doi = "10.1088/1475-7516/2014/07/057",
    journal = "JCAP",
    volume = "07",
    pages = "057",
    year = "2014"
}

@article{Porto:2013qua,
    author = "Porto, Rafael A. and Senatore, Leonardo and Zaldarriaga, Matias",
    title = "{The Lagrangian-space Effective Field Theory of Large Scale Structures}",
    eprint = "1311.2168",
    archivePrefix = "arXiv",
    primaryClass = "astro-ph.CO",
    doi = "10.1088/1475-7516/2014/05/022",
    journal = "JCAP",
    volume = "05",
    pages = "022",
    year = "2014"
}

@ARTICLE{1994ApJ...428...28M,
       author = {{Melott}, Adrian L. and {Shandarin}, Sergei F. and {Weinberg}, David H.},
        title = "{A Test of the Adhesion Approximation for Gravitational Clustering}",
      journal = {\apj},
         year = 1994,
        month = jun,
       volume = {428},
        pages = {28},
          doi = {10.1086/174216},
archivePrefix = {arXiv},
       eprint = {astro-ph/9311075},
 primaryClass = {astro-ph},
       adsurl = {https://ui.adsabs.harvard.edu/abs/1994ApJ...428...28M}
}

@article{Kofman:1991wt,
    author = "Kofman, Lev and Pogosian, Dmitri and Melott, Adrian and Shandarin, Sergei",
    title = "{Coherent structures in the universe and the adhesion model}",
    reportNumber = "CITA-91-35",
    doi = "10.1086/171517",
    journal = "Astrophys. J.",
    volume = "393",
    pages = "437--449",
    year = "1992"
}

@ARTICLE{Widrow:1993qq,
       author = "Widrow, Lawrence M. and Kaiser, Nick",
        title = "{Using the Schroedinger Equation to Simulate Collisionless Matter}",
      journal = "Astrophys. J. Lett.",
         year = 1993,
        month = oct,
       volume = {416},
        pages = {L71},
          doi = {10.1086/187073},
       adsurl = {https://ui.adsabs.harvard.edu/abs/1993ApJ...416L..71W}
}

@article{Coles:2001fw,
    author = "Coles, Peter",
    title = "{The Origin of spatial intermittency in the galaxy distribution}",
    eprint = "astro-ph/0110615",
    archivePrefix = "arXiv",
    doi = "10.1046/j.1365-8711.2002.05096.x",
    journal = "Mon. Not. Roy. Astron. Soc.",
    volume = "330",
    pages = "421",
    year = "2002"
}

@article{Coles:2002sj,
    author = "Coles, Peter and Spencer, Kate",
    title = "{A wave-mechanical approach to cosmic structure formation}",
    eprint = "astro-ph/0212433",
    archivePrefix = "arXiv",
    doi = "10.1046/j.1365-8711.2003.06529.x",
    journal = "Mon. Not. Roy. Astron. Soc.",
    volume = "342",
    pages = "176",
    year = "2003"
}

@article{EuclidTheoryWorkingGroup:2012gxx,
    author = "Amendola, Luca and others",
    collaboration = "Euclid Theory Working Group",
    title = "{Cosmology and fundamental physics with the Euclid satellite}",
    eprint = "1206.1225",
    archivePrefix = "arXiv",
    primaryClass = "astro-ph.CO",
    doi = "10.12942/lrr-2013-6",
    journal = "Living Rev. Rel.",
    volume = "16",
    pages = "6",
    year = "2013"
}

@article{Euclid:2024yrr,
    author = "Mellier, Y. and others",
    collaboration = "Euclid",
    title = "{Euclid. I. Overview of the Euclid mission}",
    eprint = "2405.13491",
    archivePrefix = "arXiv",
    primaryClass = "astro-ph.CO",
    doi = "10.1051/0004-6361/202450810",
    journal = "Astron. Astrophys.",
    volume = "697",
    pages = "A1",
    year = "2025"
}

@article{Euclid:2014mgp,
    author = "Scaramella, R. and others",
    editor = "Heavens, Alan and Starck, Jean-Luc and Krone-Martins, Alberto",
    collaboration = "Euclid",
    title = "{Euclid space mission: a cosmological challenge for the next 15 years}",
    eprint = "1501.04908",
    archivePrefix = "arXiv",
    primaryClass = "astro-ph.CO",
    doi = "10.1017/S1743921314011089",
    journal = "IAU Symp.",
    volume = "306",
    pages = "375--378",
    year = "2014"
}

@article{Maartens:2015mra,
    author = "Maartens, Roy and Abdalla, Filipe B. and Jarvis, Matt and Santos, Mario G.",
    editor = "Bourke, Tyler L. and others",
    collaboration = "SKA Cosmology SWG",
    title = "{Overview of Cosmology with the SKA}",
    eprint = "1501.04076",
    archivePrefix = "arXiv",
    primaryClass = "astro-ph.CO",
    doi = "10.22323/1.215.0016",
    journal = "PoS",
    volume = "AASKA14",
    pages = "016",
    year = "2015"
}

@article{SKA:2018ckk,
    author = "Bacon, David J. and others",
    collaboration = "SKA",
    title = "{Cosmology with Phase 1 of the Square Kilometre Array: Red Book 2018: Technical specifications and performance forecasts}",
    eprint = "1811.02743",
    archivePrefix = "arXiv",
    primaryClass = "astro-ph.CO",
    doi = "10.1017/pasa.2019.51",
    journal = "Publ. Astron. Soc. Austral.",
    volume = "37",
    pages = "e007",
    year = "2020"
}

@article{Ellis:1971pg,
    author = "Ellis, G. F. R.",
    title = "{Relativistic cosmology}",
    doi = "10.1007/s10714-009-0760-7",
    journal = "Proc. Int. Sch. Phys. Fermi",
    volume = "47",
    pages = "104--182",
    year = "1971"
}

@book{Mukhanov:2005sc,
    author = "Mukhanov, V.",
    title = "{Physical Foundations of Cosmology}",
    doi = "10.1017/CBO9780511790553",
    isbn = "978-0-521-56398-7",
    publisher = "Cambridge University Press",
    address = "Oxford",
    year = "2005"
}

@article{Short_2006a,
   title={Gravitational instability via the Schrödinger equation},
   volume={2006},
   ISSN={1475-7516},
   url={http://dx.doi.org/10.1088/1475-7516/2006/12/012},
   DOI={10.1088/1475-7516/2006/12/012},
   number={12},
   journal={Journal of Cosmology and Astroparticle Physics},
   publisher={IOP Publishing},
   author={Short, C J and Coles, P},
   year={2006},
   month=dec, pages={012–012} }

@article{Short_2006b,
   title={Wave mechanics and the adhesion approximation},
   volume={2006},
   ISSN={1475-7516},
   url={http://dx.doi.org/10.1088/1475-7516/2006/12/016},
   DOI={10.1088/1475-7516/2006/12/016},
   number={12},
   journal={Journal of Cosmology and Astroparticle Physics},
   publisher={IOP Publishing},
   author={Short, C J and Coles, P},
   year={2006},
   month=dec, pages={016–016} }

@article{Gallagher_2022,
   title={Evolution of Cosmic Voids in the Schrödinger-Poisson Formalism},
   volume={5},
   ISSN={2565-6120},
   url={http://dx.doi.org/10.21105/astro.2208.13851},
   DOI={10.21105/astro.2208.13851},
   journal={The Open Journal of Astrophysics},
   publisher={Maynooth University},
   author={Gallagher, Aoibhinn and Coles, Peter},
   year={2022},
   month=nov }

@article{Szapudi_2003,
   title={Cosmological Perturbation Theory Using the Schrödinger Equation},
   volume={583},
   ISSN={1538-4357},
   url={http://dx.doi.org/10.1086/368013},
   DOI={10.1086/368013},
   number={1},
   journal={The Astrophysical Journal},
   publisher={American Astronomical Society},
   author={Szapudi, István and Kaiser, Nick},
   year={2003},
   month=jan, pages={L1–L4} }

@article{Johnston2010,
    author = {Johnston, Rebecca and Lasenby, A. N. and Hobson, M. P.},
    title = {Cosmological fluid dynamics in the Schrödinger formalism},
    journal = {Monthly Notices of the Royal Astronomical Society},
    volume = {402},
    number = {4},
    pages = {2491-2502},
    year = {2010},
    month = {03},
    issn = {0035-8711},
    doi = {10.1111/j.1365-2966.2009.16052.x},
    url = {https://doi.org/10.1111/j.1365-2966.2009.16052.x},
    eprint = {https://academic.oup.com/mnras/article-pdf/402/4/2491/4897464/mnras0402-2491.pdf},
}

@misc{coles2002wavemechanicslargescalestructure,
      title={The Wave Mechanics of Large-scale Structure}, 
      author={Peter Coles},
      year={2002},
      eprint={astro-ph/0209576},
      archivePrefix={arXiv},
      primaryClass={astro-ph},
      url={https://arxiv.org/abs/astro-ph/0209576}, 
}

@misc{coles2025,
      title={Classical Fluid Analogies for Schr\"odinger-Newton Systems}, 
      author={Peter Coles and Aoibhinn Gallagher},
      year={2025},
      eprint={2507.08583},
      archivePrefix={arXiv},
      primaryClass={astro-ph.CO},
      url={https://arxiv.org/abs/2507.08583}, 
}

@article{Uhlemann:2014npa,
    author = "Uhlemann, Cora and Kopp, Michael and Haugg, Thomas",
    title = {{Schr{\"o}dinger method as $N$-body double and UV completion of dust}},
    eprint = "1403.5567",
    archivePrefix = "arXiv",
    primaryClass = "astro-ph.CO",
    doi = "10.1103/PhysRevD.90.023517",
    journal = "Phys. Rev. D",
    volume = "90",
    number = "2",
    pages = "023517",
    year = "2014"
}

@article{Uhlemann:2018gzz,
    author = "Uhlemann, Cora and Rampf, Cornelius and Gosenca, Mateja and Hahn, Oliver",
    title = "{Semiclassical path to cosmic large-scale structure}",
    eprint = "1812.05633",
    archivePrefix = "arXiv",
    primaryClass = "astro-ph.CO",
    doi = "10.1103/PhysRevD.99.083524",
    journal = "Phys. Rev. D",
    volume = "99",
    number = "8",
    pages = "083524",
    year = "2019"
}

@article{Gough_2022,
   title={Making (dark matter) waves: Untangling wave interference for multi-streaming dark matter},
   volume={5},
   ISSN={2565-6120},
   url={http://dx.doi.org/10.21105/astro.2206.11918},
   DOI={10.21105/astro.2206.11918},
   number={1},
   journal={The Open Journal of Astrophysics},
   publisher={Maynooth University},
   author={Gough, Alex and Uhlemann, Cora},
   year={2022},
   month=sep }

@article{Uhlemann_Kopp_2014, 
title={Beyond single-stream with the Schrödinger method}, 
volume={11}, 
DOI={10.1017/S1743921316009716}, 
number={S308}, 
journal={Proceedings of the International Astronomical Union}, 
author={Uhlemann, Cora and Kopp, Michael}, 
year={2014}, 
pages={115–118}}

@article{Madelung:1927ksh,
    author = "Madelung, E.",
    title = "{Quantentheorie in hydrodynamischer Form}",
    doi = "10.1007/BF01400372",
    journal = "Z. Phys.",
    volume = "40",
    number = "3",
    pages = "322--326",
    year = "1927"
}

@book{Griffiths_Schroeter_2018, 
place={Cambridge},
edition={3}, 
title={Introduction to Quantum Mechanics}, publisher={Cambridge University Press}, 
author={Griffiths, David J. and Schroeter, Darrell F.}, 
year={2018}
}

@book{Ellisbook,
title = "Relativistic cosmology",
author = "Ellis, \{George F. R.\} and Roy Maartens and MacCallum, \{Malcolm A. H.\}",
year = "2012",
doi = "10.1017/CBO9781139014403",
language = "English",
isbn = "9780521381154",
publisher = "Cambridge University Press",
address = "United Kingdom",
}

@article{Kofman_1995,
   title={Dynamics of gravitational instability is nonlocal},
   volume={442},
   ISSN={1538-4357},
   url={http://dx.doi.org/10.1086/175419},
   DOI={10.1086/175419},
   journal={The Astrophysical Journal},
   publisher={American Astronomical Society},
   author={Kofman, Lev and Pogosyan, Dmitry},
   year={1995},
   month=mar, pages={30} }

@article{Matarrese_1994,
   title={General relativistic dynamics of irrotational dust: Cosmological implications},
   volume={72},
   ISSN={0031-9007},
   url={http://dx.doi.org/10.1103/PhysRevLett.72.320},
   DOI={10.1103/physrevlett.72.320},
   number={3},
   journal={Physical Review Letters},
   publisher={American Physical Society (APS)},
   author={Matarrese, Sabino and Pantano, Ornella and Saez, Diego},
   year={1994},
   month=jan, pages={320–323} }

@article{Rampf_2014,
   title={Frame dragging and Eulerian frames in general relativity},
   volume={89},
   ISSN={1550-2368},
   url={http://dx.doi.org/10.1103/PhysRevD.89.063509},
   DOI={10.1103/physrevd.89.063509},
   number={6},
   journal={Physical Review D},
   publisher={American Physical Society (APS)},
   author={Rampf, Cornelius},
   year={2014},
   month=mar }

@article{Estabrook1964,
    author = {Estabrook, Frank B. and Wahlquist, Hugo D.},
    title = {Dyadic Analysis of Space-Time Congruences},
    journal = {Journal of Mathematical Physics},
    volume = {5},
    number = {11},
    pages = {1629-1644},
    year = {1964},
    month = {11},
    issn = {0022-2488},
    doi = {10.1063/1.1931200},
    url = {https://doi.org/10.1063/1.1931200},
}

@article{Nichols_2011,
   title={Visualizing spacetime curvature via frame-drag vortexes and tidal tendexes: General theory and weak-gravity applications},
   volume={84},
   ISSN={1550-2368},
   url={http://dx.doi.org/10.1103/PhysRevD.84.124014},
   DOI={10.1103/physrevd.84.124014},
   number={12},
   journal={Physical Review D},
   publisher={American Physical Society (APS)},
   author={Nichols, David A. and Owen, Robert and Zhang, Fan and Zimmerman, Aaron and Brink, Jeandrew and Chen, Yanbei and Kaplan, Jeffrey D. and Lovelace, Geoffrey and Matthews, Keith D. and Scheel, Mark A. and Thorne, Kip S.},
   year={2011},
   month=dec }

@article{PhysRevD.22.1882,
  title = {Gauge-invariant cosmological perturbations},
  author = {Bardeen, James M.},
  journal = {Phys. Rev. D},
  volume = {22},
  issue = {8},
  pages = {1882--1905},
  numpages = {0},
  year = {1980},
  month = {Oct},
  publisher = {American Physical Society},
  doi = {10.1103/PhysRevD.22.1882},
  url = {https://link.aps.org/doi/10.1103/PhysRevD.22.1882}
}

@article{Kodama:1984ziu,
    author = "Kodama, Hideo and Sasaki, Misao",
    title = "{Cosmological Perturbation Theory}",
    doi = "10.1143/PTPS.78.1",
    journal = "Prog. Theor. Phys. Suppl.",
    volume = "78",
    pages = "1--166",
    year = "1984"
}

@article{Zelrev1989,
  title = {The large-scale structure of the universe: Turbulence, intermittency, structures in a self-gravitating medium},
  author = {Shandarin, S. F. and Zeldovich, Ya. B.},
  journal = {Rev. Mod. Phys.},
  volume = {61},
  issue = {2},
  pages = {185--220},
  numpages = {0},
  year = {1989},
  month = {Apr},
  publisher = {American Physical Society},
  doi = {10.1103/RevModPhys.61.185},
  url = {https://link.aps.org/doi/10.1103/RevModPhys.61.185}
}

@article{Hui2021,
   title={Wave Dark Matter},
   volume={59},
   ISSN={1545-4282},
   url={http://dx.doi.org/10.1146/annurev-astro-120920-010024},
   DOI={10.1146/annurev-astro-120920-010024},
   number={1},
   journal={Annual Review of Astronomy and Astrophysics},
   publisher={Annual Reviews},
   author={Hui, Lam},
   year={2021},
   month=sep, pages={247–289} }

@book{Kimball:2023vxk,
    editor = "Kimball, Derek F. Jackson and van Bibber, Karl",
    title = "{The Search for Ultralight Bosonic Dark Matter}",
    doi = "10.1007/978-3-030-95852-7",
    isbn = "978-3-030-95851-0, 978-3-030-95854-1, 978-3-030-95852-7",
    publisher = "Springer",
    year = "2023"
}

@article{Ferreira_2021,
   title={Ultra-light dark matter},
   volume={29},
   ISSN={1432-0754},
   url={http://dx.doi.org/10.1007/s00159-021-00135-6},
   DOI={10.1007/s00159-021-00135-6},
   number={1},
   journal={The Astronomy and Astrophysics Review},
   publisher={Springer Science and Business Media LLC},
   author={Ferreira, Elisa G. M.},
   year={2021},
   month=sep }

@article{Brook_2022,
   title={Gravitational Stability of Vortices in Bose-Einstein Condensate Dark Matter},
   volume={5},
   ISSN={2565-6120},
   url={http://dx.doi.org/10.21105/astro.0902.0605},
   DOI={10.21105/astro.0902.0605},
   number={1},
   journal={The Open Journal of Astrophysics},
   publisher={Maynooth University},
   author={Brook, Mark N and Coles, Peter},
   year={2022},
   month=sep }

@article{PhysRevD.102.083518,
  title = {Simulating mixed fuzzy and cold dark matter},
  author = {Schwabe, Bodo and Gosenca, Mateja and Behrens, Christoph and Niemeyer, Jens C. and Easther, Richard},
  journal = {Phys. Rev. D},
  volume = {102},
  issue = {8},
  pages = {083518},
  numpages = {10},
  year = {2020},
  month = {Oct},
  publisher = {American Physical Society},
  doi = {10.1103/PhysRevD.102.083518},
  url = {https://link.aps.org/doi/10.1103/PhysRevD.102.083518}
}

@article{Hui_2017,
   title={Ultralight scalars as cosmological dark matter},
   volume={95},
   ISSN={2470-0029},
   url={http://dx.doi.org/10.1103/PhysRevD.95.043541},
   DOI={10.1103/physrevd.95.043541},
   number={4},
   journal={Physical Review D},
   publisher={American Physical Society (APS)},
   author={Hui, Lam and Ostriker, Jeremiah P. and Tremaine, Scott and Witten, Edward},
   year={2017},
   month=feb }

@ARTICLE{1965ApJ...142.1488C,
       author = {{Chandrasekhar}, S.},
        title = "{The Post-Newtonian Equations of Hydrodynamics in General Relativity}",
      journal = {\apj},
         year = 1965,
        month = nov,
       volume = {142},
        pages = {1488},
          doi = {10.1086/148432},
       adsurl = {https://ui.adsabs.harvard.edu/abs/1965ApJ...142.1488C}
}

@article{Szekeres_2000,
   title={Post-Newtonian Cosmology},
   volume={32},
   ISSN={1572-9532},
   url={http://dx.doi.org/10.1023/A:1001976317159},
   DOI={10.1023/a:1001976317159},
   number={3},
   journal={General Relativity and Gravitation},
   publisher={Springer Science and Business Media LLC},
   author={Szekeres, Peter and Rainsford, Tamath},
   year={2000},
   month=mar, pages={479–490} }

@article{Szekeres:2000ki,
    author = "Szekeres, Peter",
    title = "{Newtonian and PostNewtonian limits of relativistic cosmology}",
    doi = "10.1023/A:1001965526092",
    journal = "Gen. Rel. Grav.",
    volume = "32",
    pages = "1025--1039",
    year = "2000"
}

@article{Gressel_2019,
   title={Full-sky weak lensing: a nonlinear post-Friedmann treatment},
   volume={2019},
   ISSN={1475-7516},
   url={http://dx.doi.org/10.1088/1475-7516/2019/05/045},
   DOI={10.1088/1475-7516/2019/05/045},
   number={05},
   journal={Journal of Cosmology and Astroparticle Physics},
   publisher={IOP Publishing},
   author={Gressel, Hedda A. and Bonvin, Camille and Bruni, Marco and Bacon, David},
   year={2019},
   month=may, pages={045–045} }

@article{Rampf_2016,
   title={Lagrangian theory for cosmic structure formation with vorticity: Newtonian and post-Friedmann approximations},
   volume={94},
   ISSN={2470-0029},
   url={http://dx.doi.org/10.1103/PhysRevD.94.083515},
   DOI={10.1103/physrevd.94.083515},
   number={8},
   journal={Physical Review D},
   publisher={American Physical Society (APS)},
   author={Rampf, Cornelius and Villa, Eleonora and Bertacca, Daniele and Bruni, Marco},
   year={2016},
   month=oct }

@article{Thomas_2015,
   title={f(R) gravity on non-linear scales: the post-Friedmann expansion and the vector potential},
   volume={2015},
   ISSN={1475-7516},
   url={http://dx.doi.org/10.1088/1475-7516/2015/07/051},
   DOI={10.1088/1475-7516/2015/07/051},
   number={07},
   journal={Journal of Cosmology and Astroparticle Physics},
   publisher={IOP Publishing},
   author={Thomas, D.B. and Bruni, M. and Koyama, K. and Li, B. and Zhao, G.-B.},
   year={2015},
   month=jul, pages={051–051} }

@ARTICLE{concordance_model,
       author = {{Ostriker}, J.~P. and {Steinhardt}, Paul J.},
        title = "{The observational case for a low-density Universe with a non-zero cosmological constant}",
      journal = {\nat},
         year = 1995,
        month = oct,
       volume = {377},
       number = {6550},
        pages = {600-602},
          doi = {10.1038/377600a0},
       adsurl = {https://ui.adsabs.harvard.edu/abs/1995Natur.377..600O}
}

@article{Planck,
    author = "Aghanim, N. and others",
    collaboration = "Planck",
    title = "{Planck 2018 results. VI. Cosmological parameters}",
    eprint = "1807.06209",
    archivePrefix = "arXiv",
    primaryClass = "astro-ph.CO",
    doi = "10.1051/0004-6361/201833910",
    journal = "Astron. Astrophys.",
    volume = "641",
    pages = "A6",
    year = "2020",
    note = "[Erratum: Astron.Astrophys. 652, C4 (2021)]"
}

@article{DESI_BAO,
    author = "Abdul Karim, M. and others",
    collaboration = "DESI",
    title = "{DESI DR2 results. II. Measurements of baryon acoustic oscillations and cosmological constraints}",
    eprint = "2503.14738",
    archivePrefix = "arXiv",
    primaryClass = "astro-ph.CO",
    reportNumber = "FERMILAB-PUB-25-0169-PPD",
    doi = "10.1103/tr6y-kpc6",
    journal = "Phys. Rev. D",
    volume = "112",
    number = "8",
    pages = "083515",
    year = "2025"
}

@article{DESI_fullshape,
    author = "Adame, A. G. and others",
    collaboration = "DESI",
    title = "{DESI 2024 VII: cosmological constraints from the full-shape modeling of clustering measurements}",
    eprint = "2411.12022",
    archivePrefix = "arXiv",
    primaryClass = "astro-ph.CO",
    reportNumber = "FERMILAB-PUB-24-0854-PPD",
    doi = "10.1088/1475-7516/2025/07/028",
    journal = "JCAP",
    volume = "07",
    pages = "028",
    year = "2025"
}

@article{Type_Ia_supernovae,
   title={The Pantheon+Analysis: Cosmological Constraints},
   volume={938},
   ISSN={1538-4357},
   url={http://dx.doi.org/10.3847/1538-4357/ac8e04},
   DOI={10.3847/1538-4357/ac8e04},
   number={2},
   journal={The Astrophysical Journal},
   publisher={American Astronomical Society},
   author={Brout, Dillon and Scolnic, Dan and Popovic, Brodie and Riess, Adam G. and Carr, Anthony and Zuntz, Joe and Kessler, Rick and Davis, Tamara M. and Hinton, Samuel and Jones, David and Kenworthy, W. D’Arcy and Peterson, Erik R. and Said, Khaled and Taylor, Georgie and Ali, Noor and Armstrong, Patrick and Charvu, Pranav and Dwomoh, Arianna and Meldorf, Cole and Palmese, Antonella and Qu, Helen and Rose, Benjamin M. and Sanchez, Bruno and Stubbs, Christopher W. and Vincenzi, Maria and Wood, Charlotte M. and Brown, Peter J. and Chen, Rebecca and Chambers, Ken and Coulter, David A. and Dai, Mi and Dimitriadis, Georgios and Filippenko, Alexei V. and Foley, Ryan J. and Jha, Saurabh W. and Kelsey, Lisa and Kirshner, Robert P. and Möller, Anais and Muir, Jessie and Nadathur, Seshadri and Pan, Yen-Chen and Rest, Armin and Rojas-Bravo, Cesar and Sako, Masao and Siebert, Matthew R. and Smith, Mat and Stahl, Benjamin E. and Wiseman, Phil},
   year={2022},
   month=oct, pages={110} }

@article{Lifshitz:1945du,
    author = "Lifshitz, E.",
    title = "{Republication of: On the gravitational stability of the expanding universe}",
    doi = "10.1007/s10714-016-2165-8",
    journal = "J. Phys. (USSR)",
    volume = "10",
    number = "2",
    pages = "116",
    year = "1946"
}

@ARTICLE{Blumenthal1984Natur,
       author = {{Blumenthal}, G.~R. and {Faber}, S.~M. and {Primack}, J.~R. and {Rees}, M.~J.},
        title = "{Formation of galaxies and large-scale structure with cold dark matter.}",
      journal = {\nat},
         year = 1984,
        month = Oct,
       volume = {311},
        pages = {517-525},
          doi = {10.1038/311517a0},
       adsurl = {https://ui.adsabs.harvard.edu/abs/1984Natur.311..517B}
}

@ARTICLE{Peebles1982ApJ,
       author = {{Peebles}, P.J.E.},
        title = "{Large-scale background temperature and mass fluctuations due to scale-invariant primeval perturbations}",
      journal = {ApJL},
         year = 1982,
        month = dec,
       volume = {263},
        pages = {L1-L5},
          doi = {10.1086/183911},
       adsurl = {https://ui.adsabs.harvard.edu/abs/1982ApJ...263L...1P}
}

@article{DESI:2016fyo,
author = {{DESI Collaboration} and {Aghamousa}, Amir and {Aguilar}, Jessica and {Ahlen}, Steve and {Alam}, Shadab and {Allen}, Lori E. and {Allende Prieto}, Carlos and {Annis}, James and {Bailey}, Stephen and {Balland}, Christophe and {Ballester}, Otger and {Baltay}, Charles and {Beaufore}, Lucas and {Bebek}, Chris and {Beers}, Timothy C. and {Bell}, Eric F. and {Bernal}, Jos{\'e} Luis and {Besuner}, Robert and {Beutler}, Florian and {Blake}, Chris and {Bleuler}, Hannes and {Blomqvist}, Michael and {Blum}, Robert and {Bolton}, Adam S. and {Briceno}, Cesar and {Brooks}, David and {Brownstein}, Joel R. and {Buckley-Geer}, Elizabeth and {Burden}, Angela and {Burtin}, Etienne and {Busca}, Nicolas G. and {Cahn}, Robert N. and {Cai}, Yan-Chuan and {Cardiel-Sas}, Laia and {Carlberg}, Raymond G. and {Carton}, Pierre-Henri and {Casas}, Ricard and {Castander}, Francisco J. and {Cervantes-Cota}, Jorge L. and {Claybaugh}, Todd M. and {Close}, Madeline and {Coker}, Carl T. and {Cole}, Shaun and {Comparat}, Johan and {Cooper}, Andrew P. and {Cousinou}, M.-C. and {Crocce}, Martin and {Cuby}, Jean-Gabriel and {Cunningham}, Daniel P. and {Davis}, Tamara M. and {Dawson}, Kyle S. and {de la Macorra}, Axel and {De Vicente}, Juan and {Delubac}, Timoth{\'e}e and {Derwent}, Mark and {Dey}, Arjun and {Dhungana}, Govinda and {Ding}, Zhejie and {Doel}, Peter and {Duan}, Yutong T. and {Ealet}, Anne and {Edelstein}, Jerry and {Eftekharzadeh}, Sarah and {Eisenstein}, Daniel J. and {Elliott}, Ann and {Escoffier}, St{\'e}phanie and {Evatt}, Matthew and {Fagrelius}, Parker and {Fan}, Xiaohui and {Fanning}, Kevin and {Farahi}, Arya and {Farihi}, Jay and {Favole}, Ginevra and {Feng}, Yu and {Fernandez}, Enrique and {Findlay}, Joseph R. and {Finkbeiner}, Douglas P. and {Fitzpatrick}, Michael J. and {Flaugher}, Brenna and {Flender}, Samuel and {Font-Ribera}, Andreu and {Forero-Romero}, Jaime E. and {Fosalba}, Pablo and {Frenk}, Carlos S. and {Fumagalli}, Michele and {Gaensicke}, Boris T. and {Gallo}, Giuseppe and {Garcia-Bellido}, Juan and {Gaztanaga}, Enrique and {Pietro Gentile Fusillo}, Nicola and {Gerard}, Terry and {Gershkovich}, Irena and {Giannantonio}, Tommaso and {Gillet}, Denis and {Gonzalez-de-Rivera}, Guillermo and {Gonzalez-Perez}, Violeta and {Gott}, Shelby and {Graur}, Or and {Gutierrez}, Gaston and {Guy}, Julien and {Habib}, Salman and {Heetderks}, Henry and {Heetderks}, Ian and {Heitmann}, Katrin and {Hellwing}, Wojciech A. and {Herrera}, David A. and {Ho}, Shirley and {Holland}, Stephen and {Honscheid}, Klaus and {Huff}, Eric and {Hutchinson}, Timothy A. and {Huterer}, Dragan and {Hwang}, Ho Seong and {Illa Laguna}, Joseph Maria and {Ishikawa}, Yuzo and {Jacobs}, Dianna and {Jeffrey}, Niall and {Jelinsky}, Patrick and {Jennings}, Elise and {Jiang}, Linhua and {Jimenez}, Jorge and {Johnson}, Jennifer and {Joyce}, Richard and {Jullo}, Eric and {Juneau}, St{\'e}phanie and {Kama}, Sami and {Karcher}, Armin and {Karkar}, Sonia and {Kehoe}, Robert and {Kennamer}, Noble and {Kent}, Stephen and {Kilbinger}, Martin and {Kim}, Alex G. and {Kirkby}, David and {Kisner}, Theodore and {Kitanidis}, Ellie and {Kneib}, Jean-Paul and {Koposov}, Sergey and {Kovacs}, Eve and {Koyama}, Kazuya and {Kremin}, Anthony and {Kron}, Richard and {Kronig}, Luzius and {Kueter-Young}, Andrea and {Lacey}, Cedric G. and {Lafever}, Robin and {Lahav}, Ofer and {Lambert}, Andrew and {Lampton}, Michael and {Landriau}, Martin and {Lang}, Dustin and {Lauer}, Tod R. and {Le Goff}, Jean-Marc and {Le Guillou}, Laurent and {Le Van Suu}, Auguste and {Lee}, Jae Hyeon and {Lee}, Su-Jeong and {Leitner}, Daniela and {Lesser}, Michael and {Levi}, Michael E. and {L'Huillier}, Benjamin and {Li}, Baojiu and {Liang}, Ming and {Lin}, Huan and {Linder}, Eric and {Loebman}, Sarah R. and {Luki{\'c}}, Zarija and {Ma}, Jun and {MacCrann}, Niall and {Magneville}, Christophe and {Makarem}, Laleh and {Manera}, Marc and {Manser}, Christopher J. and {Marshall}, Robert and {Martini}, Paul and {Massey}, Richard and {Matheson}, Thomas and {McCauley}, Jeremy and {McDonald}, Patrick and {McGreer}, Ian D. and {Meisner}, Aaron and {Metcalfe}, Nigel and {Miller}, Timothy N. and {Miquel}, Ramon and {Moustakas}, John and {Myers}, Adam and {Naik}, Milind and {Newman}, Jeffrey A. and {Nichol}, Robert C. and {Nicola}, Andrina and {Nicolati da Costa}, Luiz and {Nie}, Jundan and {Niz}, Gustavo and {Norberg}, Peder and {Nord}, Brian and {Norman}, Dara and {Nugent}, Peter and {O'Brien}, Thomas and {Oh}, Minji and {Olsen}, Knut A.~G.},
        title = "{The DESI Experiment Part I: Science,Targeting, and Survey Design}",
      journal = {arXiv e-prints},
         year = 2016,
        month = oct,
          eid = {arXiv:1611.00036},
        pages = {arXiv:1611.00036},
          doi = {10.48550/arXiv.1611.00036},
archivePrefix = {arXiv},
       eprint = {1611.00036},
 primaryClass = {astro-ph.IM},
       adsurl = {https://ui.adsabs.harvard.edu/abs/2016arXiv161100036D}
}

@article{LSSTDarkEnergyScience:2012kar,
 author = {{LSST Dark Energy Science Collaboration}},
        title = "{Large Synoptic Survey Telescope: Dark Energy Science Collaboration}",
      journal = {arXiv e-prints},
         year = 2012,
        month = nov,
          eid = {arXiv:1211.0310},
        pages = {arXiv:1211.0310},
          doi = {10.48550/arXiv.1211.0310},
archivePrefix = {arXiv},
       eprint = {1211.0310},
 primaryClass = {astro-ph.CO},
       adsurl = {https://ui.adsabs.harvard.edu/abs/2012arXiv1211.0310L}
}

@article{Roman,
author = {{Observations Time Allocation Committee}, Roman and {Community Survey Definition Committees}, Core},
        title = "{Roman Observations Time Allocation Committee: Final Report and Recommendations}",
      journal = {arXiv e-prints},
         year = 2025,
        month = may,
          eid = {arXiv:2505.10574},
        pages = {arXiv:2505.10574},
          doi = {10.48550/arXiv.2505.10574},
archivePrefix = {arXiv},
       eprint = {2505.10574},
 primaryClass = {astro-ph.IM},
       adsurl = {https://ui.adsabs.harvard.edu/abs/2025arXiv250510574O}
}

@article{Wands2008,
    author = "Malik, Karim A. and Wands, David",
    title = "{Cosmological perturbations}",
    eprint = "0809.4944",
    archivePrefix = "arXiv",
    primaryClass = "astro-ph",
    doi = "10.1016/j.physrep.2009.03.001",
    journal = "Phys. Rept.",
    volume = "475",
    pages = "1--51",
    year = "2009"
}

@article{PecceiQuinn,
  title = {$\mathrm{CP}$ Conservation in the Presence of Pseudoparticles},
  author = {Peccei, R. D. and Quinn, Helen R.},
  journal = {Phys. Rev. Lett.},
  volume = {38},
  issue = {25},
  pages = {1440--1443},
  numpages = {0},
  year = {1977},
  month = {Jun},
  publisher = {American Physical Society},
  doi = {10.1103/PhysRevLett.38.1440},
  url = {https://link.aps.org/doi/10.1103/PhysRevLett.38.1440}
}

@article{Preskill:1982cy,
    author = "Preskill, John and Wise, Mark B. and Wilczek, Frank",
    editor = "Srednicki, M. A.",
    title = "{Cosmology of the Invisible Axion}",
    reportNumber = "HUTP-82-A048, NSF-ITP-82-103",
    doi = "10.1016/0370-2693(83)90637-8",
    journal = "Phys. Lett. B",
    volume = "120",
    pages = "127--132",
    year = "1983"
}

@article{Abbott:1982af,
    author = "Abbott, L. F. and Sikivie, P.",
    editor = "Srednicki, M. A.",
    title = "{A Cosmological Bound on the Invisible Axion}",
    reportNumber = "PRINT-82-0695 (BRANDEIS)",
    doi = "10.1016/0370-2693(83)90638-X",
    journal = "Phys. Lett. B",
    volume = "120",
    pages = "133--136",
    year = "1983"
}

@article{Dine:1982ah,
    author = "Dine, Michael and Fischler, Willy",
    editor = "Srednicki, M. A.",
    title = "{The Not So Harmless Axion}",
    reportNumber = "UPR-0201T",
    doi = "10.1016/0370-2693(83)90639-1",
    journal = "Phys. Lett. B",
    volume = "120",
    pages = "137--141",
    year = "1983"
}

@article{70PhysRevLett.61.2175,
  title = {Approximation Scheme for Constructing a Clumpy Universe in General Relativity},
  author = {Futamase, Toshifumi},
  journal = {Phys. Rev. Lett.},
  volume = {61},
  issue = {19},
  pages = {2175--2178},
  numpages = {0},
  year = {1988},
  month = {Nov},
  publisher = {American Physical Society},
  doi = {10.1103/PhysRevLett.61.2175},
  url = {https://link.aps.org/doi/10.1103/PhysRevLett.61.2175}
}

@article{72,
    author = {Tomita, Kenji},
    title = {Post-Newtonian Equations of Motion in Flat Cosmological Models},
    journal = {Progress of Theoretical Physics},
    volume = {85},
    number = {5},
    pages = {1041-1047},
    year = {1991},
    month = {05},
    issn = {0033-068X},
    doi = {10.1143/ptp/85.5.1041},
    url = {https://doi.org/10.1143/ptp/85.5.1041},
    eprint = {https://academic.oup.com/ptp/article-pdf/85/5/1041/5454753/85-5-1041.pdf},
}

@article{73PhysRevD.53.681,
  title = {Averaging of a locally inhomogeneous realistic universe},
  author = {Futamase, Toshifumi},
  journal = {Phys. Rev. D},
  volume = {53},
  issue = {2},
  pages = {681--689},
  numpages = {0},
  year = {1996},
  month = {Jan},
  publisher = {American Physical Society},
  doi = {10.1103/PhysRevD.53.681},
  url = {https://link.aps.org/doi/10.1103/PhysRevD.53.681}
}

@article{75Takada1997APL,
  title={A post-Newtonian Lagrangian perturbation approach to large-scale structure formation},
  author={Masahiro Takada and Toshifumi Futamase},
  journal={Monthly Notices of the Royal Astronomical Society},
  year={1997},
  volume={306},
  pages={64-88},
  url={https://api.semanticscholar.org/CorpusID:14767104}
}

@article{76Matarrese_1996,
   title={Post-Newtonian cosmological dynamics in Lagrangian coordinates},
   volume={283},
   ISSN={1365-2966},
   url={http://dx.doi.org/10.1093/mnras/283.2.400},
   DOI={10.1093/mnras/283.2.400},
   number={2},
   journal={Monthly Notices of the Royal Astronomical Society},
   publisher={Oxford University Press (OUP)},
   author={Matarrese, S. and Terranova, D.},
   year={1996},
   month=nov, pages={400–418} }

@article{Green:2010qy,
    author = "Green, Stephen R. and Wald, Robert M.",
    title = "{A new framework for analyzing the effects of small scale inhomogeneities in cosmology}",
    eprint = "1011.4920",
    archivePrefix = "arXiv",
    primaryClass = "gr-qc",
    doi = "10.1103/PhysRevD.83.084020",
    journal = "Phys. Rev. D",
    volume = "83",
    pages = "084020",
    year = "2011"
}

@article{Green:2011wc,
    author = "Green, Stephen R. and Wald, Robert M.",
    title = "{Newtonian and Relativistic Cosmologies}",
    eprint = "1111.2997",
    archivePrefix = "arXiv",
    primaryClass = "gr-qc",
    doi = "10.1103/PhysRevD.85.063512",
    journal = "Phys. Rev. D",
    volume = "85",
    pages = "063512",
    year = "2012"
}

@article{Carbone:2004iv,
    author = "Carbone, Carmelita and Matarrese, Sabino",
    title = "{A Unified treatment of cosmological perturbations from super-horizon to small scales}",
    eprint = "astro-ph/0407611",
    archivePrefix = "arXiv",
    reportNumber = "DFPD-04-A-18",
    doi = "10.1103/PhysRevD.71.043508",
    journal = "Phys. Rev. D",
    volume = "71",
    pages = "043508",
    year = "2005"
}

@article{48Chisari:2011iq,
    author = "Chisari, Nora Elisa and Zaldarriaga, Matias",
    title = "{Connection between Newtonian simulations and general relativity}",
    eprint = "1101.3555",
    archivePrefix = "arXiv",
    primaryClass = "astro-ph.CO",
    doi = "10.1103/PhysRevD.84.089901",
    journal = "Phys. Rev. D",
    volume = "83",
    pages = "123505",
    year = "2011",
    note = "[Erratum: Phys.Rev.D 84, 089901 (2011)]"
}

@article{50Bruni:2011ta,
    author = "Bruni, Marco and Crittenden, Robert and Koyama, Kazuya and Maartens, Roy and Pitrou, Cyril and Wands, David",
    title = "{Disentangling non-Gaussianity, bias and GR effects in the galaxy distribution}",
    eprint = "1106.3999",
    archivePrefix = "arXiv",
    primaryClass = "astro-ph.CO",
    doi = "10.1103/PhysRevD.85.041301",
    journal = "Phys. Rev. D",
    volume = "85",
    pages = "041301",
    year = "2012"
}

@ARTICLE{52Bartolo2010,
       author = {{Bartolo}, Nicola and {Matarrese}, Sabino and {Pantano}, Ornella and {Riotto}, Antonio},
        title = "{Second-order matter perturbations in a {\ensuremath{\Lambda}}CDM cosmology and non-Gaussianity}",
      journal = {Classical and Quantum Gravity},
         year = 2010,
        month = jun,
       volume = {27},
       number = {12},
          eid = {124009},
        pages = {124009},
          doi = {10.1088/0264-9381/27/12/124009},
archivePrefix = {arXiv},
       eprint = {1002.3759},
 primaryClass = {astro-ph.CO},
       adsurl = {https://ui.adsabs.harvard.edu/abs/2010CQGra..27l4009B}
}

@article{53Bruni:2013qta,
    author = "Bruni, Marco and Hidalgo, Juan Carlos and Meures, Nikolai and Wands, David",
    title = "{Non-Gaussian Initial Conditions in {\ensuremath{\Lambda}}CDM: Newtonian, Relativistic, and Primordial Contributions}",
    eprint = "1307.1478",
    archivePrefix = "arXiv",
    primaryClass = "astro-ph.CO",
    doi = "10.1088/0004-637X/785/1/2",
    journal = "Astrophys. J.",
    volume = "785",
    pages = "2",
    year = "2014"
}

@article{54Bruni:2014xma,
    author = "Bruni, Marco and Hidalgo, Juan Carlos and Wands, David",
    title = "{Einstein's signature in cosmological large-scale structure}",
    eprint = "1405.7006",
    archivePrefix = "arXiv",
    primaryClass = "astro-ph.CO",
    doi = "10.1088/2041-8205/794/1/L11",
    journal = "Astrophys. J. Lett.",
    volume = "794",
    number = "1",
    pages = "L11",
    year = "2014"
}

@article{55Kopp:2013tqa,
    author = "Kopp, Michael and Uhlemann, Cora and Haugg, Thomas",
    title = "{Newton to Einstein {\textemdash} dust to dust}",
    eprint = "1312.3638",
    archivePrefix = "arXiv",
    primaryClass = "astro-ph.CO",
    doi = "10.1088/1475-7516/2014/03/018",
    journal = "JCAP",
    volume = "03",
    pages = "018",
    year = "2014"
}

@article{57Green:2014aga,
    author = "Green, Stephen R. and Wald, Robert M.",
    title = "{How well is our universe described by an FLRW model?}",
    eprint = "1407.8084",
    archivePrefix = "arXiv",
    primaryClass = "gr-qc",
    doi = "10.1088/0264-9381/31/23/234003",
    journal = "Class. Quant. Grav.",
    volume = "31",
    pages = "234003",
    year = "2014"
}

@article{58Villa:2014foa,
    author = "Villa, Eleonora and Verde, Licia and Matarrese, Sabino",
    title = "{General relativistic corrections and non-Gaussianity in large scale structure}",
    eprint = "1409.4738",
    archivePrefix = "arXiv",
    primaryClass = "astro-ph.CO",
    doi = "10.1088/0264-9381/31/23/234005",
    journal = "Class. Quant. Grav.",
    volume = "31",
    number = "23",
    pages = "234005",
    year = "2014"
}

@article{59Rampf:2014xpt,
    author = "Rampf, Cornelius and Rigopoulos, Gerasimos and Valkenburg, Wessel",
    title = "{A Relativistic view on large scale N-body simulations}",
    eprint = "1409.6549",
    archivePrefix = "arXiv",
    primaryClass = "astro-ph.CO",
    doi = "10.1088/0264-9381/31/23/234004",
    journal = "Class. Quant. Grav.",
    volume = "31",
    pages = "234004",
    year = "2014"
}

@article{Frame_draggingBruni:2013mua,
    author = "Bruni, Marco and Thomas, Daniel B. and Wands, David",
    title = "{Computing General Relativistic effects from Newtonian N-body simulations: Frame dragging in the post-Friedmann approach}",
    eprint = "1306.1562",
    archivePrefix = "arXiv",
    primaryClass = "astro-ph.CO",
    doi = "10.1103/PhysRevD.89.044010",
    journal = "Phys. Rev. D",
    volume = "89",
    number = "4",
    pages = "044010",
    year = "2014"
}

@article{Yoo:2014sfa,
    author = "Yoo, Jaiyul and Zaldarriaga, Matias",
    title = "{Beyond the Linear-Order Relativistic Effect in Galaxy Clustering: Second-Order Gauge-Invariant Formalism}",
    eprint = "1406.4140",
    archivePrefix = "arXiv",
    primaryClass = "astro-ph.CO",
    doi = "10.1103/PhysRevD.90.023513",
    journal = "Phys. Rev. D",
    volume = "90",
    number = "2",
    pages = "023513",
    year = "2014"
}

@article{Bertschinger:1994nc,
    author = "Bertschinger, Edmund and Hamilton, A. J. S.",
    title = "{Lagrangian evolution of the Weyl tensor}",
    eprint = "astro-ph/9403016",
    archivePrefix = "arXiv",
    reportNumber = "MIT-CSR-94-06",
    doi = "10.1086/174787",
    journal = "Astrophys. J.",
    volume = "435",
    pages = "1",
    year = "1994"
}

@article{Ma:1995ey,
    author = "Ma, Chung-Pei and Bertschinger, Edmund",
    title = "{Cosmological perturbation theory in the synchronous and conformal Newtonian gauges}",
    eprint = "astro-ph/9506072",
    archivePrefix = "arXiv",
    doi = "10.1086/176550",
    journal = "Astrophys. J.",
    volume = "455",
    pages = "7--25",
    year = "1995"
}

@article{Widrow:1996eq,
    author = "Widrow, Lawrence M.",
    title = "{Modeling collisionless matter in general relativity: A New numerical technique}",
    eprint = "astro-ph/9607124",
    archivePrefix = "arXiv",
    doi = "10.1103/PhysRevD.55.5997",
    journal = "Phys. Rev. D",
    volume = "55",
    pages = "5997--6001",
    year = "1997"
}

@article{Bardeen1980,
  title = {Gauge-invariant cosmological perturbations},
  author = {Bardeen, James M.},
  journal = {Phys. Rev. D},
  volume = {22},
  issue = {8},
  pages = {1882--1905},
  numpages = {0},
  year = {1980},
  month = {Oct},
  publisher = {American Physical Society},
  doi = {10.1103/PhysRevD.22.1882},
  url = {https://link.aps.org/doi/10.1103/PhysRevD.22.1882}
}

@article{Ehlers:1961xww,
    author = "Ehlers, J.",
    title = "{Contributions to the relativistic mechanics of continuous media}",
    doi = "10.1007/BF00759031",
    journal = "Abh. Akad. Wiss. Lit. Mainz. Nat. Kl.",
    volume = "11",
    pages = "793--837",
    year = "1961"
}

@book{Misner:1973prb,
    author = "Misner, Charles W. and Thorne, K. S. and Wheeler, J. A.",
    title = "{Gravitation}",
    isbn = "978-0-7167-0344-0, 978-0-691-17779-3",
    publisher = "W. H. Freeman",
    address = "San Francisco",
    year = "1973"
}

@ARTICLE{Bardeen1970_ZAMO,
       author = {{Bardeen}, James M.},
        title = "{A Variational Principle for Rotating Stars in General Relativity}",
      journal = {\apj},
         year = 1970,
        month = oct,
       volume = {162},
        pages = {71},
          doi = {10.1086/150635},
       adsurl = {https://ui.adsabs.harvard.edu/abs/1970ApJ...162...71B}
}

@ARTICLE{Bardeen1972_LNRF,
       author = {{Bardeen}, James M. and {Press}, William H. and {Teukolsky}, Saul A.},
        title = "{Rotating Black Holes: Locally Nonrotating Frames, Energy Extraction, and Scalar Synchrotron Radiation}",
      journal = {\apj},
         year = 1972,
        month = dec,
       volume = {178},
        pages = {347-370},
          doi = {10.1086/151796},
       adsurl = {https://ui.adsabs.harvard.edu/abs/1972ApJ...178..347B}
}

@article{Beordo:2025cpw,
 author = {{Beordo}, William and {Bruni}, Marco and {Barrera-Hinojosa}, Cristian and {Crosta}, Mariateresa},
        title = "{The frame-dragging vector potential on galaxy scales from Dark-Matter-only Newtonian N-body simulations}",
      journal = {Monthly Notices of the Royal Astronomical Society},
         year = 2026,
        month = may,
       volume = {548},
       number = {2},
          eid = {stag489},
        pages = {stag489},
          doi = {10.1093/mnras/stag489},
archivePrefix = {arXiv},
       eprint = {2512.08703},
 primaryClass = {astro-ph.CO},
       adsurl = {https://ui.adsabs.harvard.edu/abs/2026MNRAS.548ag489B}
}

@article{Lense:1918zz,
    author = "Lense, Josef and Thirring, Hans",
    title = "{Ueber den Einfluss der Eigenrotation der Zentralkoerper auf die Bewegung der Planeten und Monde nach der Einsteinschen Gravitationstheorie}",
    journal = "Phys. Z.",
    volume = "19",
    pages = "156--163",
    year = "1918"
}

@article{Thomas:2015kua,
    author = "Thomas, Daniel B. and Bruni, Marco and Wands, David",
    title = "{The fully non-linear post-Friedmann frame-dragging vector potential: Magnitude and time evolution from N-body simulations}",
    eprint = "1501.00799",
    archivePrefix = "arXiv",
    primaryClass = "astro-ph.CO",
    doi = "10.1093/mnras/stv1390",
    journal = "Mon. Not. Roy. Astron. Soc.",
    volume = "452",
    number = "2",
    pages = "1727--1742",
    year = "2015"
}

\end{document}